\documentclass[preprint2]{aastex}

\usepackage{graphicx}
\usepackage{natbib}
\usepackage{txfonts}
\usepackage{color}
\usepackage{wrapfig}
\usepackage{amssymb}
\usepackage{color}
\usepackage{lscape}

\shorttitle{Detection of a hot core in the LMC}
\shortauthors{Shimonishi et al.}

\begin{document}


\title{Detection of a hot molecular core in the Large Magellanic Cloud with ALMA} 


\author{Takashi Shimonishi\altaffilmark{1,2,\bigstar}}
\author{Takashi Onaka\altaffilmark{3}}
\author{Akiko Kawamura\altaffilmark{4}}
\and
\author{Yuri Aikawa\altaffilmark{5}}



\altaffiltext{1}{Frontier Research Institute for Interdisciplinary Sciences, Tohoku University, Aramakiazaaoba 6-3, Aoba-ku, Sendai, Miyagi, 980-8578, Japan}
\altaffiltext{2}{Astronomical Institute, Tohoku University, Aramakiazaaoba 6-3, Aoba-ku, Sendai, Miyagi, 980-8578, Japan}
\altaffiltext{3}{Department of Astronomy, Graduate School of Science, The University of Tokyo, 7-3-1 Hongo, Bunkyo-ku, Tokyo 113-0033, Japan}
\altaffiltext{4}{National Astronomical Observatory of Japan, 2-21-1 Osawa, Mitaka, Tokyo, 181-8588, Japan}
\altaffiltext{5}{Center for Computational Sciences, The University of Tsukuba, 1-1-1, Tennodai, Tsukuba, Ibaraki 305-8577, Japan}
\altaffiltext{$\bigstar$}{shimonishi@astr.tohoku.ac.jp}


\begin{abstract} 
We report the first detection of a hot molecular core outside our Galaxy based on radio observations with ALMA toward a high-mass young stellar object (YSO) in a nearby low metallicity galaxy, the Large Magellanic Cloud (LMC). 
Molecular emission lines of CO, C$^{17}$O, HCO$^{+}$, H$^{13}$CO$^{+}$, H$_2$CO, NO, SiO, H$_2$CS, $^{33}$SO, $^{32}$SO$_2$, $^{34}$SO$_2$, and $^{33}$SO$_2$ are detected from a compact region ($\sim$0.1 pc) associated with a high-mass YSO, ST11. 
The temperature of molecular gas is estimated to be higher than 100 K based on rotation diagram analysis of SO$_2$ and $^{34}$SO$_2$ lines. 
The compact source size, warm gas temperature, high density, and rich molecular lines around a high-mass protostar suggest that ST11 is associated with a hot molecular core. 
We find that the molecular abundances of the LMC hot core are significantly different from those of Galactic hot cores. 
The abundances of CH$_3$OH, H$_2$CO, and HNCO are remarkably lower compared with Galactic hot cores by at least 1--3 orders of magnitude. 
We suggest that these abundances are characterized by the deficiency of molecules whose formation requires the hydrogenation of CO on grain surfaces. 
In contrast, NO shows a high abundance in ST11 despite the notably low abundance of nitrogen in the LMC. 
A multitude of SO$_2$ and its isotopologue line detections in ST11 imply that SO$_2$ can be a key molecular tracer of hot core chemistry in metal-poor environments. 
Furthermore, we find molecular outflows around the hot core, which is the second detection of an extragalactic protostellar outflow. 
In this paper, we discuss physical and chemical characteristics of a hot molecular core in the low metallicity environment. 
\end{abstract}


\section{Introduction} \label{sec_introduction}
Because cosmic metallicity is increasing in time with the evolution of our universe, interstellar chemistry in low metallicity environments is crucial to understand chemical processes in the past universe. 
For this purpose, observations of chemically-rich objects in nearby low metallicity galaxies and comparative studies with Galactic counterparts play an important role. 

Hot molecular cores are one of the early stages of high-mass star formation and they play a key role in the formation and evolution of complex molecules in space. 
In terms of physical properties, hot cores are defined as having small source size ($\leq$0.1 pc), high density ($\geq$10$^6$ cm$^{-3}$), and warm gas/dust temperature ($\geq$100 K) \citep[e.g.,][]{Kur00,vdT04}. 
Chemistry of hot cores is characterized by sublimation of ice mantles, which accumulated in the course of star formation. 
In cold molecular clouds and prestellar cores, gaseous molecules and atoms are frozen onto dust grains and hydrogenated. 
As the core is heated by star-formation activities, reaction among heavy species become active on grain surfaces to form larger molecules. 
In addition, sublimated molecules, such as CH$_3$OH and NH$_3$, are subject to further gas-phase reactions \citep[e.g.,][]{Gar06,Gar08b,Her09}. 
As a result, hot cores show a wealth of molecular spectral lines in infrared and radio wavelengths. 
Thus detailed studies of chemical properties of hot cores are crucial to understand complex chemical processes triggered by star formation. 

The Large Magellanic Cloud (LMC) is an excellent target to study interstellar and circumstellar chemistry in different metallicity environments owing to its proximity \citep[49.97 $\pm$ 1.11 kpc,][]{Pie13} and low metallicity \citep[about one third of the solar neighborhood,][]{Wes90}. 
The low dust content in the galaxy results in harsh radiation environment, and thus photoprocessing of interstellar medium should be more effective in the LMC than in our Galaxy \citep{Isr86}. 
Furthermore, according to gamma-ray observations, the cosmic-ray density in the LMC is estimated to be lower than the Galactic typical values by a factor of four \citep{Abd10}. 
It is therefore highly anticipated that these environmental differences should affect chemical processes, and hot cores in the LMC should provide us key information to understand chemistry, particularly those of complex molecules, in low metallicity environments. 
However, so far, observations of hot cores have been limited to Galactic sources due to lack of spatial resolution and sensitivity of radio telescopes. 

Most of radio studies on chemical compositions of molecular gas in the LMC have been performed with single-dish telescopes. 
Early studies by the SEST 15 m telescope conducted multiline observations toward \ion{H}{2} regions in the LMC and detected molecular species as large as CH$_3$OH, C$_3$H$_2$ and SO$_2$ \citep{Joh94,Chin97,Hei99,Wan09}. 
Submillimeter observations of relatively dense molecular gas in star-forming regions in the LMC are also reported \citep{Par14,Par16}. 
\citet{Nis15} recently conducted deep and unbiased spectral line surveys in the 3 mm window toward a number of molecular clouds in the LMC using the Mopra telescope. 
They reported the low abundances of nitrogen-bearing molecules, the deficiency of CH$_3$OH, and the high abundance of C$_2$H in the LMC compared with Galactic molecular clouds. 

Characteristic interstellar chemistry in the LMC is also suggested from previous infrared observations of ices around embedded young stellar objects (YSOs) in the LMC. 
\citet{ST,ST10} reported that the CO$_2$/H$_2$O ice ratio of high-mass YSOs in the LMC is systematically higher than those of Galactic high-mass YSOs based on infrared observations with the \textit{AKARI} satellite. 
Recently, \citet{ST16} reported that the CH$_3$OH ice around high-mass YSOs in the LMC is less abundant compared with Galactic counterparts based on infrared observations with the Very Large Telescope. 
The authors suggest that warm ice chemistry (grain surface reactions at a relatively high dust temperature) is responsible for the observed characteristics of ice chemical compositions in the LMC. 
Furthermore, detailed studies of the 15.2 $\mu$m CO$_2$ ice band toward LMC high-mass YSOs with \textit{Spitzer} suggest a higher degree of thermal processing of ices in the LMC than in our Galaxy \citep{Oli09,Sea11}. 
Since gas-grain chemistry is believed to play an important role in hot cores, the characteristic ice chemistry in the LMC would imply diverse hot core chemistry in extragalactic environments according to their metallicities. 

High-spatial resolution interferometry observations toward star-forming regions in the LMC have been reported with the Australia Telescope Compact Array \citep[e.g.,][]{Won06,Ott08,Sea12,And14} and recently with the Atacama Large Millimeter/submillimeter Array (ALMA) \citep[e.g.,][]{Ind13,Fuk15}. 
These observations resolve star-forming regions in the LMC down to parsec- or sub-parsec-scale and investigate physical properties of dense molecular gas. 
However, chemical properties of warm and dense molecular gas associated with a single high-mass YSO in the LMC still remain to be investigated. 

In this paper, we report the detection of a hot molecular core in the LMC based on submillimeter interferometric observations with ALMA. 
In $\S$ \ref{sec_obs} we describe the details of observations and data reduction conducted in this work. 
Obtained images and spectra of molecular line emission and continuum are presented in $\S$ \ref{sec_results}. 
Analysis of spectral line data and derivation of physical quantities of molecular gas and dust are described in $\S$ \ref{sec_analysis}. 
Physical and chemical properties of the observed source are discussed in $\S$ \ref{sec_discussions}. 
Finally, conclusions of this paper are summarized in $\S$ \ref{sec_summary}. 

\begin{figure*}[!]
\begin{center}
\includegraphics[width=12.2cm]{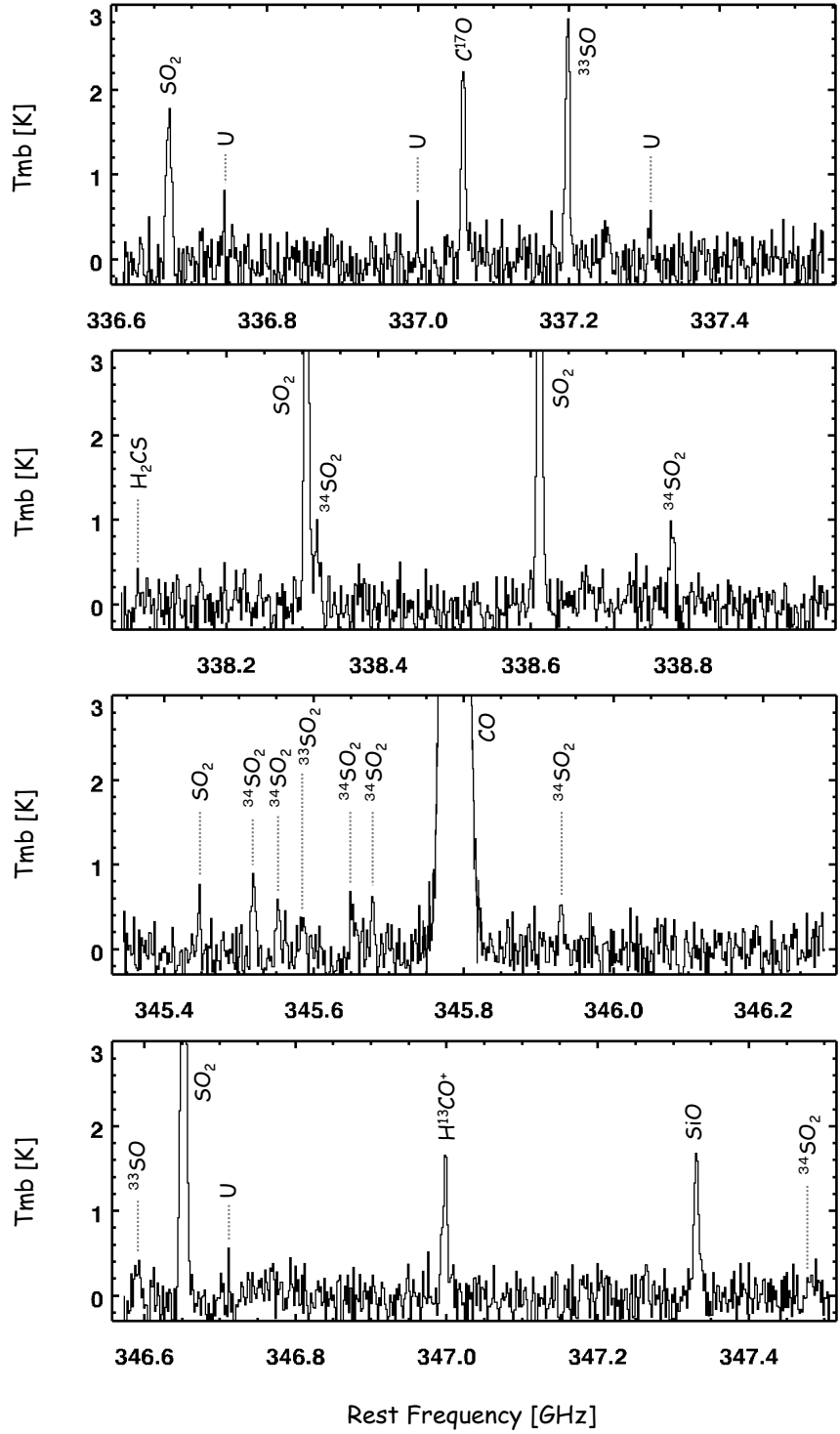}
\caption{ALMA band 7 spectra of ST11 extracted from the central region with a radius of 0.5$\arcsec$. 
Detected emission lines are labeled. 
Tentative detections are indicated by ``\textit{?}" and unidentified lines are by ``\textit{U}". 
The adopted source velocity is 250.5 km/s. 
}
\label{spec1}
\end{center}
\end{figure*}


\begin{figure*}[!t]
\begin{center}
\includegraphics[width=12.2cm]{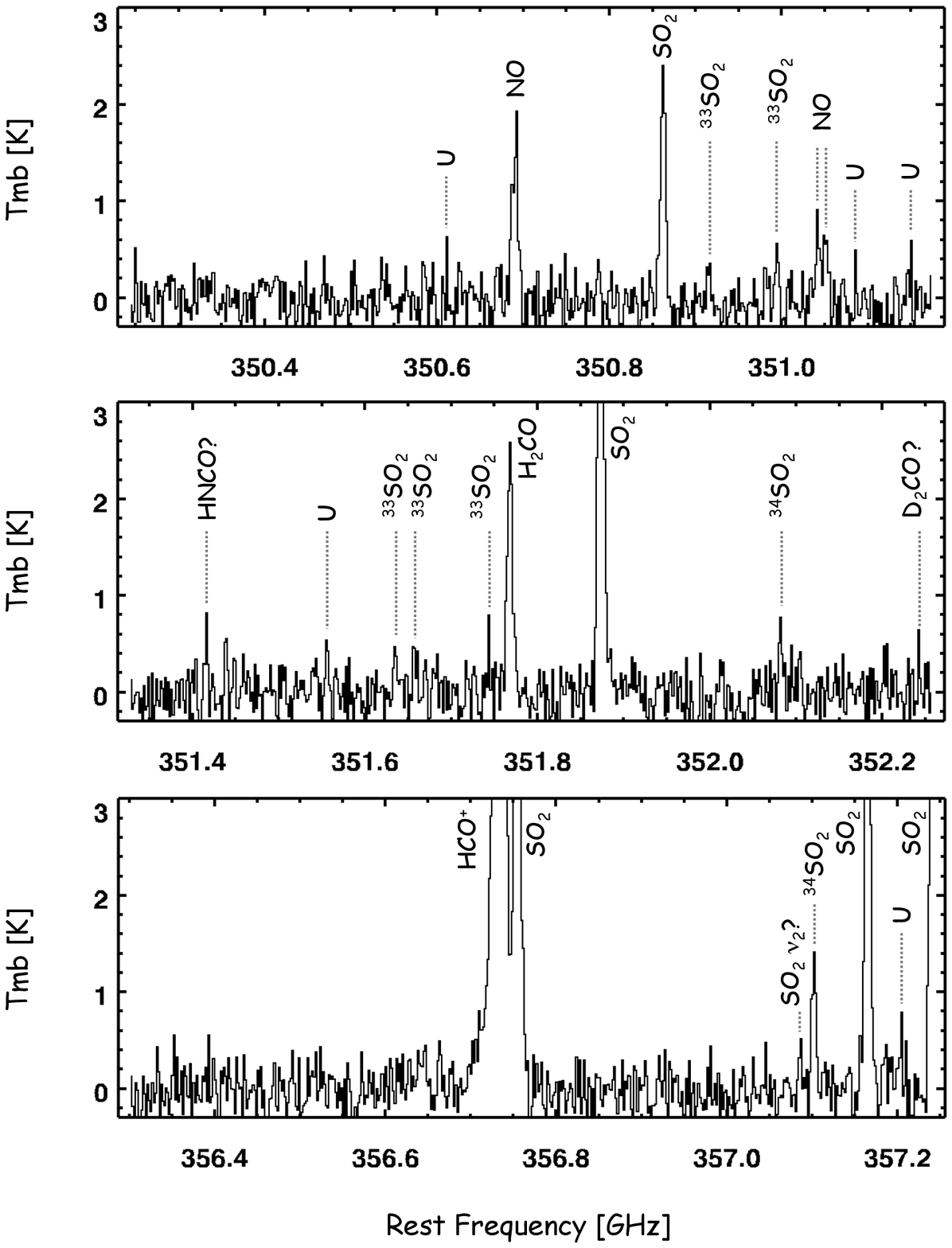}
\caption{{\it Continued} }
\label{spec2}
\end{center}
\end{figure*}

\section{Observations and data reduction} \label{sec_obs}
\subsection{Target} \label{sec_target} 
The target of the present observations is a high-mass YSO, 2MASS J05264658-6848469 or ST11 (hereafter ST11), located in the LMC. 
The source is spectroscopically identified to be a high-mass YSO in previous infrared studies \citep{Sea09,ST10}. 
Detailed YSO properties of ST11 are revisited in this work and discussed in $\S$ \ref{sec_LM}.

\subsection{Observations} \label{sec_observations}
Observations were carried out with ALMA between November 2013 and February 2014 as a part of the Cycle 1 high priority program 2012.1.01108.S (PI T. Shimonishi). 
The telescopes were pointed to RA = 05$^\mathrm{h}$26$^\mathrm{m}$46$\fs$63 and Dec = -68$^\circ$48$\arcmin$47$\farcs$10 (J2000), which is a position of ST11 measured by infrared data \citep{ST10}. 
The compact configurations C32-3 and partially C32-2 were used for our observations. 
The target object was observed in Band 7 with seven spectral bands covering 336.60--337.55, 338.05-339.00, 345.35--346.30, 346.55--347.50, 350.25--351.20, 351.30--352.25, and 356.30--357.25 GHz in the rest-frame frequency. 
The observed lines and the integration times are presented in Table \ref{tab_lines}. 
The primary beam has a full-width at half-maximum (FWHM) of 18--19$\arcsec$ at these frequencies, which corresponds to the field-of-view (FoV) of imaging data. 
The maximum recoverable angular scale is about 7$\arcsec$. 
The velocity resolution of the original data is 0.4 km/s for the spectral bands and 26 km/s for the continuum band.

\subsection{Data reduction} \label{sec_reduction}
Raw interferometric data is processed using the \textit{Common Astronomy Software Applications} (CASA) package. 
The calibration is done by CASA 4.1.0 and imaging as well as spectral extraction are done by CASA 4.3.0. 
The flux calibrator is J0519-454 and the phase calibrators are J0601-7036 and J0635-7516. 
The synthesized beam size in the 338 GHz region is approximately 0.5$\arcsec$ $\times$ 0.5$\arcsec$, which corresponds to 0.12 pc at the assumed distance to the LMC \citep[49.97 kpc,][]{Pie13}. 
The primary beam correction is done by the impbcor task in CASA, but the correction have little effect on the extracted spectra since the target source is located at the center of FoV and very compact. 

The spectra as well as continuum flux are extracted from the circular region with a diameter of 0.5$\arcsec$ centered at RA = 05$^\mathrm{h}$26$^\mathrm{m}$46$\fs$60 and Dec = -68$^\circ$48$\arcmin$47$\farcs$03 (J2000). 
This region corresponds to the peak position of the 359 GHz (840 $\mu$m) continuum emission of ST11 measured in this study, and the diameter corresponds to the beam size. 
The continuum emission is subtracted from the spectral data using the uvcontsub task in CASA. 
Several channels are concatenated during the clean process to increase S/N and the channel spacing of the reduced data is 1.5 km/s (1.73 MHz) except for CO(3--2). 
For the CO(3--2) line, the spectral region is not binned since the line is sufficiently strong, and the channel spacing is 0.4 km/s (0.46 MHz).

\section{Results} \label{sec_results}
\subsection{Observed spectra} \label{sec_spectra}
Figures \ref{spec1}--\ref{spec2} show the spectra extracted from the 0.5$\arcsec$ diameter region centered at ST11. 
In the figures the sky frequency is converted to the rest frequency using the LSR velocity of 250.5 km/s, which is the typical radial velocity of molecular lines detected toward the source. 
Spectral lines are identified with the aid of the Cologne Database for Molecular Spectroscopy\footnote{https://www.astro.uni-koeln.de/cdms} \citep[CDMS,][]{Mul01,Mul05} and the molecular database of the Jet Propulsion Laboratory\footnote{http://spec.jpl.nasa.gov} \citep[JPL,][]{Pic98}. 
Molecular emission lines due to CO, C$^{17}$O, HCO$^{+}$, H$^{13}$CO$^{+}$, H$_2$CO, NO, SiO, H$_2$CS, $^{33}$SO, $^{32}$SO$_2$, $^{34}$SO$_2$, and $^{33}$SO$_2$ are detected from a compact region associated with ST11. 
A number of high excitation lines ($E_u$ $>$100 K) are detected for SO$_2$ and its isotopologues. 
The emission lines of the above molecules in the 345 GHz band are for the first time detected toward the LMC source except for CO and HCO$^{+}$. 
Emission lines of CH$_3$OH, HNCO, CS, HC$_3$N, and complex organic molecules are not detected. 
Unidentified lines are labeled in the figures, but some of them may be spurious signals. 

\begin{figure*}[t]
\begin{center}
\includegraphics[width=15.0cm]{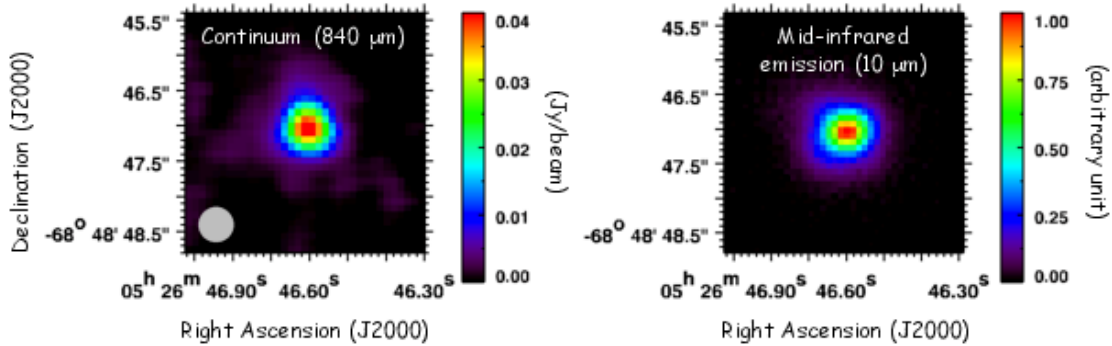}
\caption{Flux distribution of the ALMA 840 $\mu$m continuum data tracing cold dust (left) and the Gemini/T-ReCS mid-infrared 10 $\mu$m image tracing warm dust (right). 
}
\label{image_continuum}
\end{center}
\end{figure*}


\begin{figure*}[t]
\begin{center}
\includegraphics[width=15.6cm]{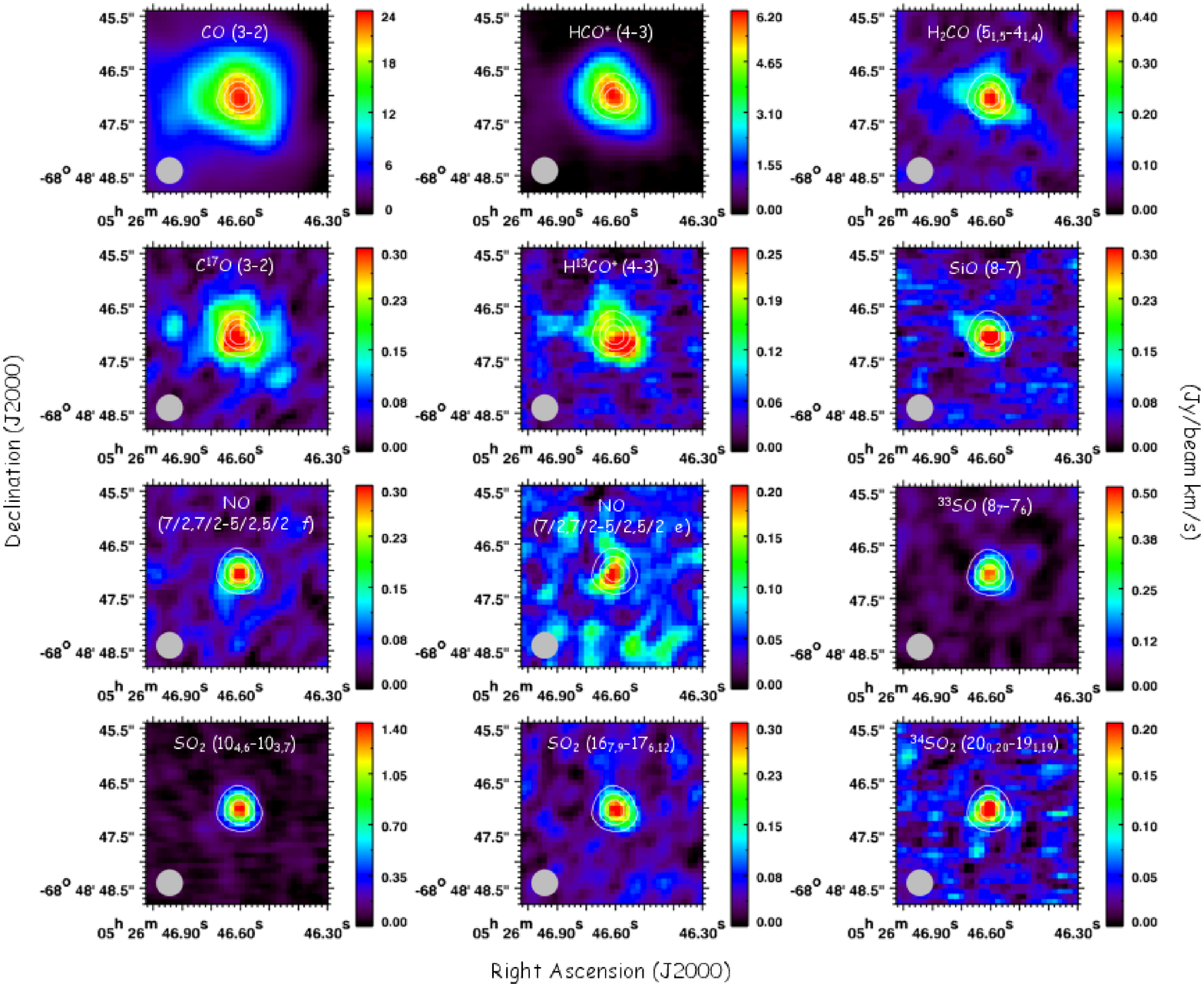}
\caption{
Integrated intensity distributions of CO, C$^{17}$O, HCO$^+$, H$^{13}$CO$^+$, H$_2$CO, SiO, NO, $^{33}$SO, SO$_2$, and $^{34}$SO$_2$ lines. 
Contours represent the distribution of the 840 $\mu$m continuum, and the contour levels are 25 $\%$, 50 $\%$ and 75 $\%$ of the peak flux. 
The synthesized beam size (0.5$\arcsec$, 0.12 pc at the LMC) is shown by the gray filled circle in each panel. 
}
\label{image_Others}
\end{center}
\end{figure*}

\subsection{Synthesized images} \label{sec_images}
Figures \ref{image_continuum}--\ref{image_Others} show synthesized images of continuum and molecular emission lines observed toward ST11. 
For SO$_2$ and $^{34}$SO$_2$, we only show representative lines because distributions of emission are similar in different transitions. 
The lines of SO$_2$ ($\nu$$_2$ = 1), $^{33}$SO$_2$, and H$_2$CS are not included in the figure because they are too weak to visualize the brightness distribution. 
The images are constructed by integrating each spectrum in the velocity range where the emission line is seen, typically between 240 km/s and 260 km/s. 
For CO, the spectrum is integrated between 230 km/s and 285 km/s because the line is very broad. 

A high spatial resolution mid-infrared image of ST11 is also shown in Figure \ref{image_continuum}. 
The image is obtained by T-ReCS at the Gemini South telescope (Program ID: S10B-120, PI: T. Shimonishi) and a broad-band filter in the N-band (7.70--12.97 $\mu$m, centered at 10.36 $\mu$m) is used for the observation. 
The ALMA 840 $\mu$m continuum traces distribution of cold dust, while the mid-infrared image traces warm dust. 

The source is compact in general and peak positions of each emission line coincide with the region of dust continuum emission. 
We estimate a source size (Gaussian FWHM, $\theta$) for continuum and relatively strong emission lines by a two-dimensional Gaussian fit. 
The estimated $\theta$ in the 840 $\mu$m continuum image is about 0.6$\arcsec$, which is close to the beam size. 
The mid-infrared emission shows $\theta$ = 0.5$\arcsec$, which is as compact as the 840 $\mu$m continuum. 
The molecular lines of NO, SiO, $^{33}$SO, SO$_2$, and $^{34}$SO$_2$ typically have $\theta$ $\sim$0.5$\arcsec$, which is indistinguishable from the beam size. 
Since the distribution is as compact as the beam size, these emissions are considered to be a point source with the present spatial resolution. 
The H$^{13}$CO$^+$ and H$_2$CO lines are slightly extended compared with the above lines and have $\theta$ $\sim$0.8$\arcsec$. 
The lines of CO, C$^{17}$O and HCO$^+$ are more extended and $\theta$ is 1.1$\arcsec$--1.4$\arcsec$. 
Because the source size is sufficiently smaller than the maximum recoverable angular scale of $\sim$7$\arcsec$, the emission from ST11 is almost recovered by the present interferometric observations.

\section{Analysis} \label{sec_analysis} 
\begin{figure*}[!]
\begin{center}
\includegraphics[width=14.4cm]{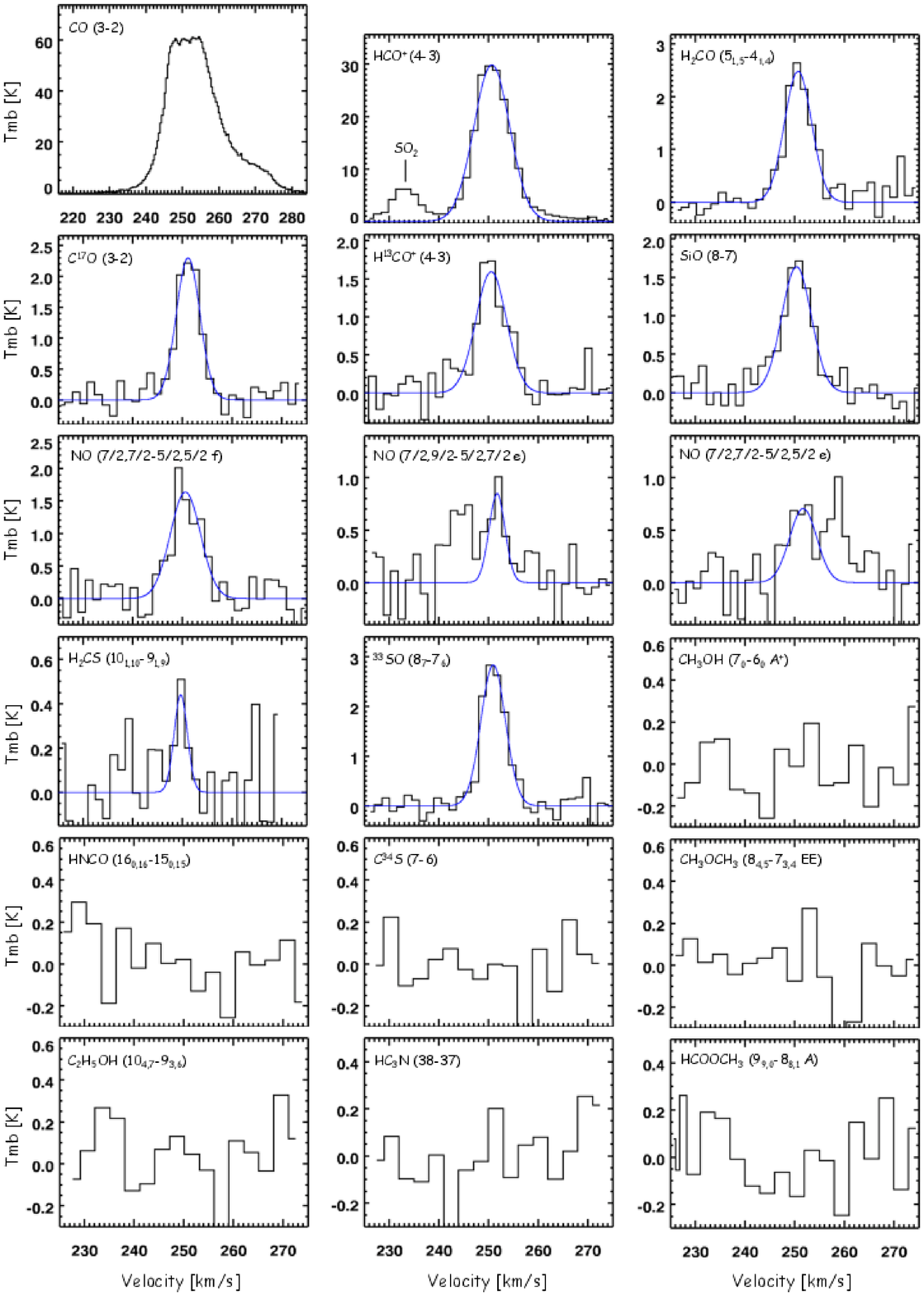}
\caption{
Spectra of CO, C$^{17}$O, HCO$^+$, H$^{13}$CO$^+$, H$_2$CO, SiO, NO, H$_2$CS, and $^{33}$SO emission lines extracted from the 0.5$\arcsec$ diameter region centered at ST11. 
The blue lines represent Gaussian profiles fitted to the observed spectra. 
The spectral regions of important non-detection lines including CH$_3$OH, HNCO, C$^{34}$S, CH$_3$OCH$_3$, C$_2$H$_5$OH, HC$_3$N, and HCOOCH$_3$ are also shown. 
}
\label{line_Others}
\end{center}
\end{figure*}


\begin{figure*}[!t]
\begin{center}
\includegraphics[width=14.4cm]{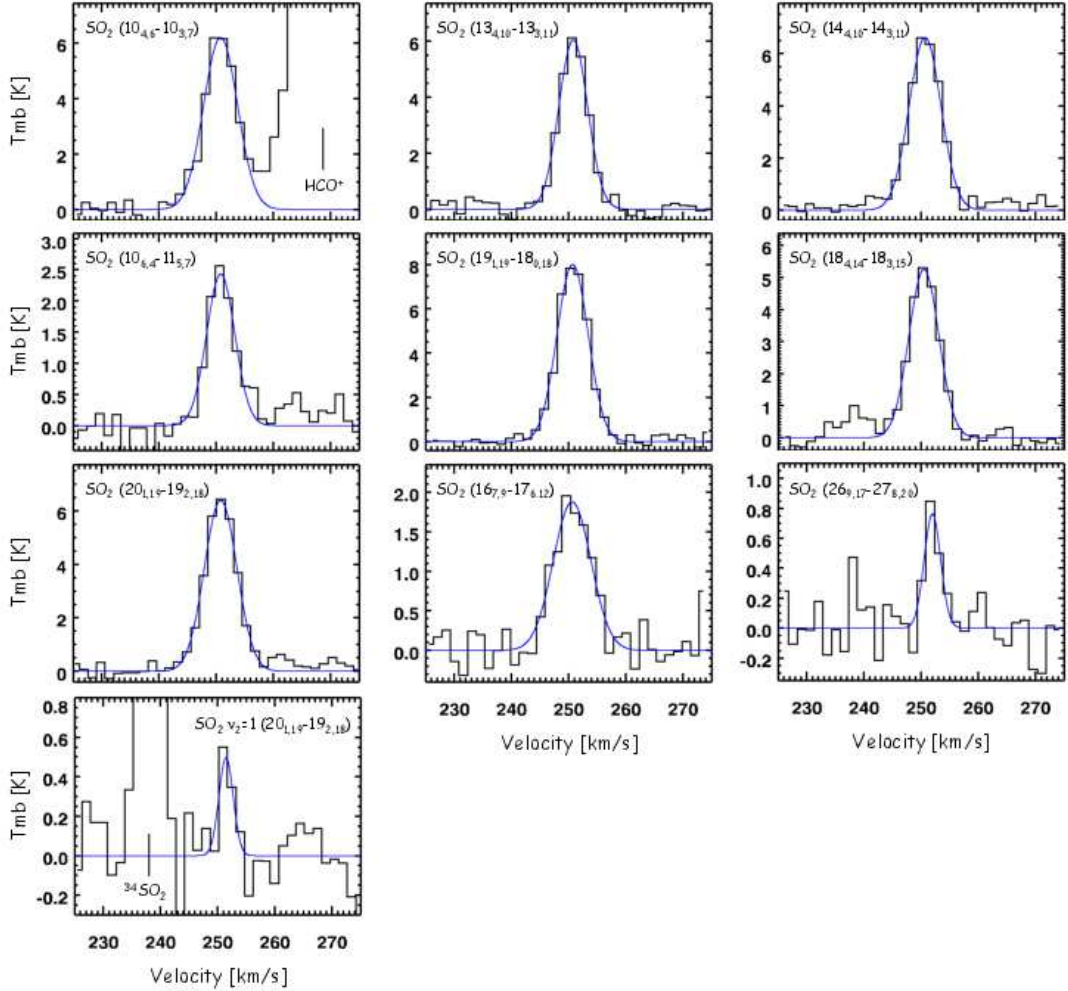}
\caption{Spectra of SO$_2$ emission lines extracted from the 0.5$\arcsec$ diameter region centered at ST11. 
The blue lines represent a Gaussian profile fitted to the observed spectra. 
The spectra are sorted in ascending order of the upper state energy (The emission line with the lowest upper state energy is shown in the upper left panel and that with the highest energy is in the lower right panel). 
The bottom panel is for the SO$_2$ ($\nu$$_2$ = 1) line. 
}
\label{line_SO2}
\end{center}
\end{figure*}


\begin{figure*}[!t]
\begin{center}
\includegraphics[width=14.4cm]{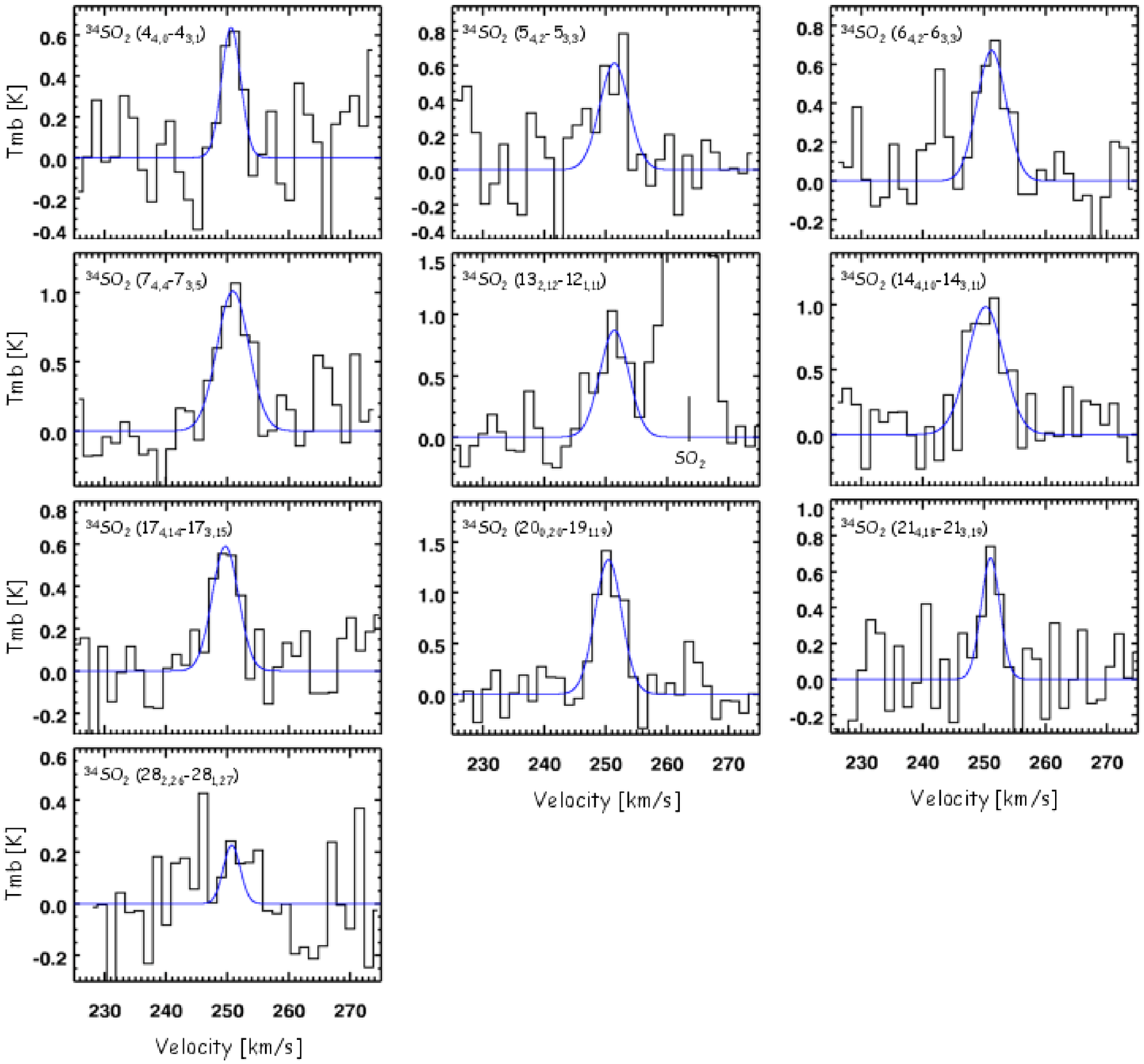}
\caption{Spectra of $^{34}$SO$_2$ emission lines observed toward ST11 as in Figure \ref{line_SO2}. 
The spectra are sorted in ascending order of the upper state energy from the upper left to the lower right. 
}
\label{line_34SO2}
\end{center}
\end{figure*}


\begin{figure*}[!t]
\begin{center}
\includegraphics[width=14.4cm]{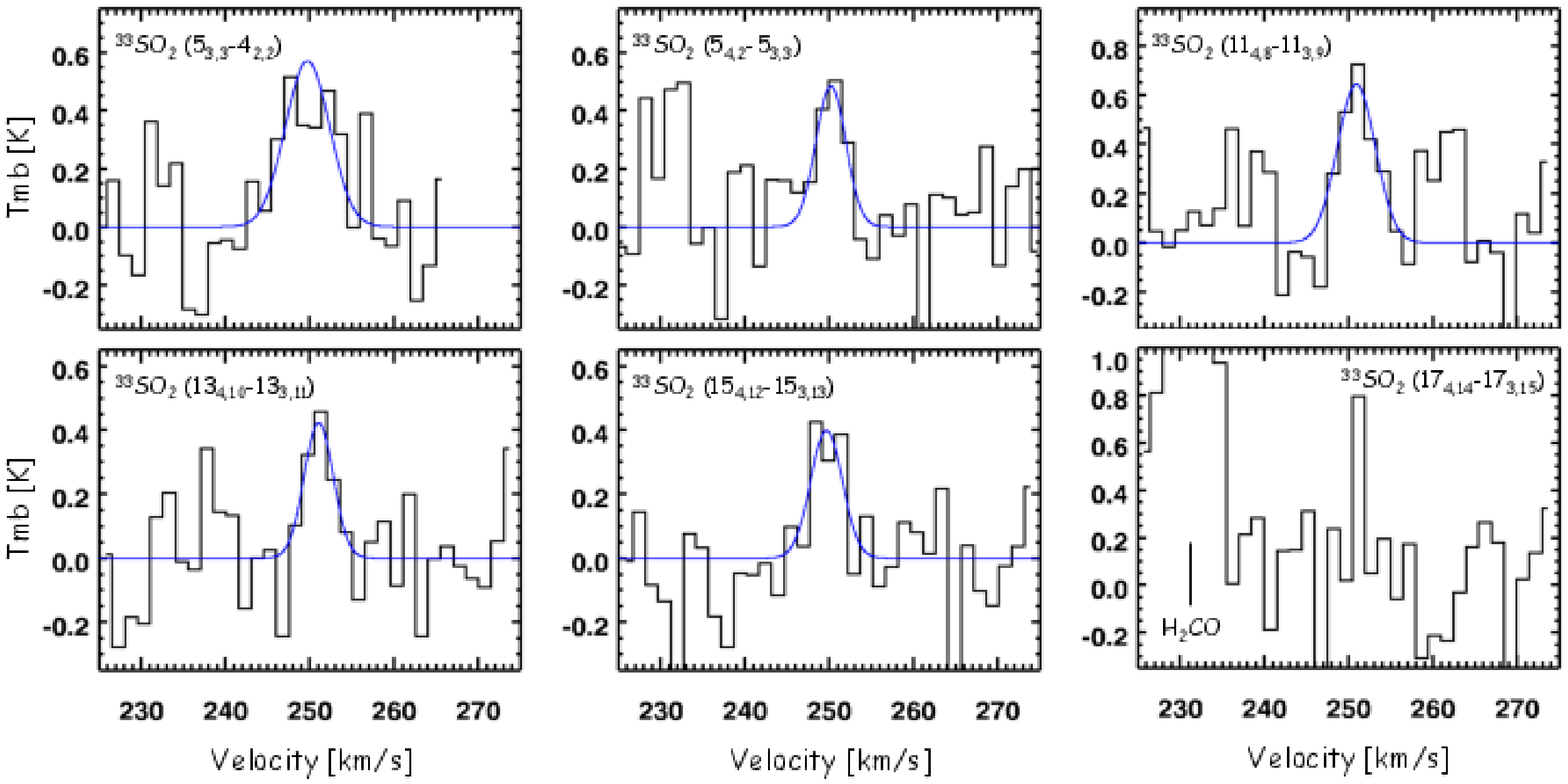}
\caption{Spectra of $^{33}$SO$_2$ emission lines observed toward ST11 as in Figure \ref{line_SO2}. 
The spectra are sorted in ascending order of the upper state energy from the upper left to the lower right. 
The 17$_{4,14}$--17$_{3,15}$ line is a tentative detection. 
}
\label{line_33SO2}
\end{center}
\end{figure*}

\subsection{Spectral fitting} \label{sec_fitting} 
The line parameters are measured by fitting a single Gaussian profile to the observed lines. 
For the NO lines at 351.0435 GHz and 351.0515 GHz, we fit a double Gaussian because they are partially blended. 
In some cases we subtract a local baseline, which is estimated from adjacent line-free regions to correct for weak baseline ripples. 
For the SO$_2$(10$_{4,6}$--10$_{3,7}$) line, we subtracted the HCO$^+$(4--3) line upon fitting since they are partly blended. 
In general, good fits are obtained with Gaussian profiles except for the CO(3--2) line, which deviates from a Gaussian. 
We estimate a peak main-beam brightness temperature, a FWHM, a LSR velocity, and an integrated intensity for each line on the basis of the fitting. 
For CO, instead, we estimate the peak brightness temperature and the FWHM by visual inspection, and the integrated intensity is estimated by integrating the spectrum in the velocity range between 230 km/s and 285 km/s. 
The spectra and the results of the gaussian fitting are shown Figures \ref{line_Others}--\ref{line_33SO2}. 
The figures also show the spectral regions of several important non-detection lines. 
The measured line parameters are summarized in Table \ref{tab_lines}. 

Multiple hyperfine components sometime exist within fitted profile according to the spectroscopic catalogues (CDMS or JPL), but these are not resolved due to the low spectral resolution of the present spectra. 
When the blended lines have comparable upper state energies, we split the measured flux according to the their $S\mu^{2}$ values and the upper state degeneracy (see $\S$ \ref{sec_rd} for definition of $S$ and $\mu$). 
Using this method, we estimate the line parameters of the strongest hyperfine line, which is used in the subsequent analysis of column densities and rotational temperatures.

\subsection{Rotation diagram analysis for SO$_2$, $^{34}$SO$_2$ and $^{33}$SO$_2$} \label{sec_rd}
Since we detect multiple emission lines with different excitation energies for SO$_2$, $^{34}$SO$_2$ and $^{33}$SO$_2$, we perform the rotation diagram analysis assuming an optically thin condition and an local thermodynamic equilibrium (LTE). 
A column density of molecules in the upper energy level, $N_{u}^{thin}$, is derived by the following equation for optically thin lines \citep[e.g.,][]{Sut95,Gol99}, 
\begin{equation}
\frac{ N_{u}^{thin} }{ g_{u} } = \frac{ 3 k \int T_{mb} dV }{ 8 \pi^{3} \nu S \mu^{2} }, \label{Eq1}
\end{equation}
where $g_{u}$ is the degeneracy of the upper level, $k$ is the Boltzmann constant,  $\int T_{mb} dV$ is the integrated intensity as estimated from the observations, $\nu$ is the transition frequency, $S$ is the line strength, and $\mu$ is the dipole moment. 
Under the LTE condition, the total column density, $N_{total}$, is given by 
\begin{equation} 
\frac{ N_{u}^{thin} }{ g_{u} } = \frac{ N_{total} }{ Q(T_{rot}) } e^{-E_{u}/kT_{rot}}, \label{Eq2}
\end{equation}
where $Q(T_{rot})$ is the partition function, $T_{rot}$ is the rotational temperature, and $E_{u}$ is the upper state energy. 
This equation is rearranged as follows. 
\begin{equation}
\log \left(\frac{ N_{u}^{thin} }{ g_{u} } \right) = - \left(\frac {\log e}{T_{rot}} \right) \left(\frac{E_{u}}{k} \right) + \log \left(\frac{N_{total}}{Q(T_{rot})} \right) \label{Eq3}
\end{equation}

When $N_{u}^{thin}/g_{u}$ is plotted against $E_{u}/k$ and data points are fitted by a straight-line, the slope and the intercept correspond to $T_{rot}$ and $N_{total}$, respectively. 
Thus we can simultaneously determine the rotational temperature and the total column density. 
All the spectroscopic parameters required in the above analysis are extracted from the CDMS or the JPL database. 
For the partition function, we interpolate the data given in the databases and estimate the appropriate $Q(T_{rot})$ at the derived rotational temperature. 

\begin{figure*}[!]
\begin{center}
\includegraphics[width=14.2cm]{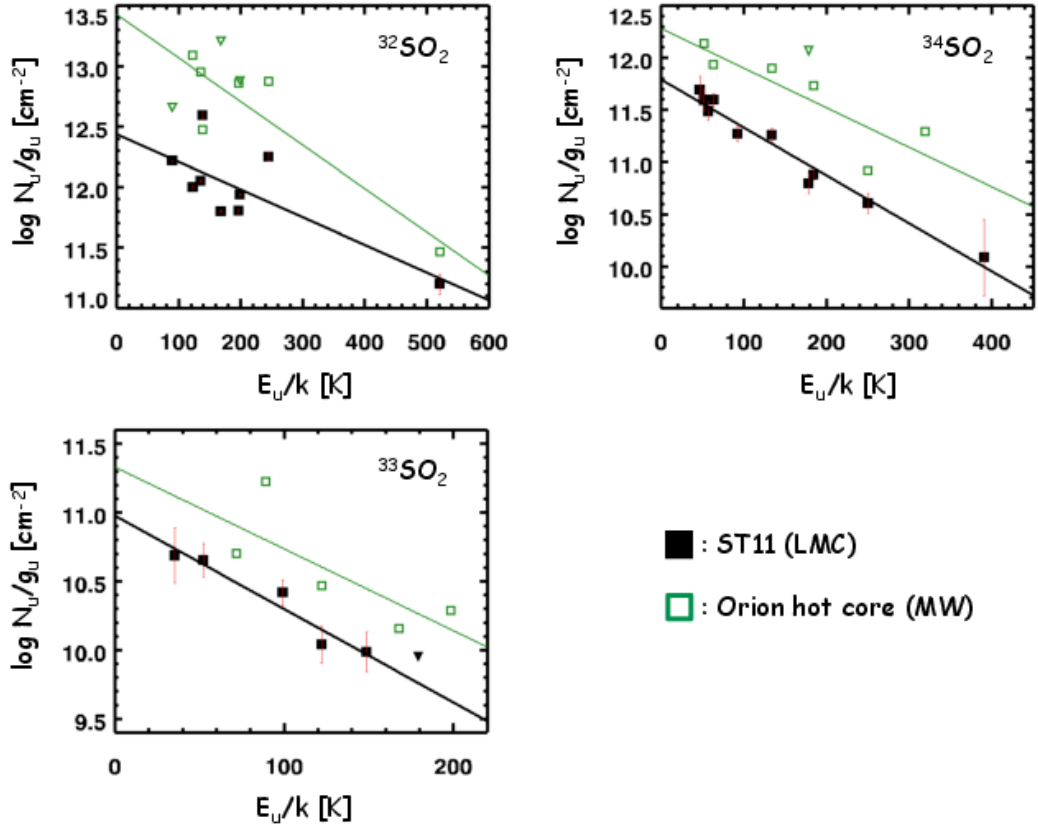}
\caption{Rotation diagram analysis of $^{32}$SO$_2$ (upper left), $^{34}$SO$_2$ (upper right), and $^{33}$SO$_2$ (lower left) for ST11. 
The filled squares (black) are for ST11 and the open squares (green) are for the Orion hot core. 
The downward triangles represent upper limits. 
The straight-lines fitted to the ST11 and the Orion data points are shown by the black and green solid lines, respectively. 
The derived rotational temperatures are shown in Table \ref{tab_T}. 
}
\label{rot_diag}
\end{center}
\end{figure*}

The constructed rotation diagrams for SO$_2$, $^{34}$SO$_2$ and $^{33}$SO$_2$ are shown in Figure \ref{rot_diag} and the derived temperatures and column densities are summarized in Table \ref{tab_N}. 
Uncertainties in the table are of 2 $\sigma$ level and do not include systematic errors due to spectroscopic parameters extracted from the CDMS and JPL databases. 

For a comparison purpose, we also perform the rotation diagram analysis for the Orion hot core using the lines with the similar spectroscopic properties with those used in the analysis of ST11. 
The Orion data is obtained with the 20$\arcsec$ beam size ($\sim$0.04 pc at the Orion region) and adopted from \citet{Schi97}. 
The rotational temperatures of the Orion hot core derived here are somewhat lower than temperatures derived in \citet{Schi97} using complete spectral line survey data in 325--360 GHz. 
This is because high excitation lines used in our comparative analysis are fewer than those used in \citet{Schi97} to allow for a fair comparison. 
The above results are shown in Figure \ref{rot_diag} and Table \ref{tab_T} together with the results of ST11. 
The table also shows rotational temperatures of SO$_2$ and $^{34}$SO$_2$ measured for W3 (H$_2$O) and G34.3+0.15, which are obtained from the literature \citep{Hel97, Mac96}.

\subsection{Column densities of other molecules} \label{sec_N_others}
Column densities of other molecular species than SO$_2$ and its isotopologues are derived by solving Eq. \ref{Eq2} for $N_{total}$ under assumption of the LTE and optically thin condition. 
We here assume that the observed molecular species are located in the same region as SO$_2$ and its isotopologues, and thus have the similar rotational temperatures; $T_{rot}$ is assumed to be 100 K, which is roughly the average rotational temperature of SO$_2$, $^{34}$SO$_2$ and $^{33}$SO$_2$. 

The similar spatial distributions of emission and the similar radial velocities of SO$_2$ and other molecules support the validity of this assumption, but more rigorous excitation analysis with further spectral data is absolutely necessary in the future. 

NO shows multiple transitions with the same upper state energy in the 350--351 GHz region. 
We estimate the column density for each transition, and the average value is adopted as the final NO column density. 
The scatter of column densities estimated from different lines is less than 20 $\%$. 

We also estimate upper limits on column densities of important non-detection lines such as CH$_3$OH, HNCO, C$^{34}$S and some complex molecules (see $\S\ref{sec_X_mol}$ for details of individual lines). 
Upper limits on peak brightness temperatures are measured using the observed spectra after appropriate channel binning, and then we assume the FWHM of 6 km/s to estimate the upper limits on integrated intensities. 
This FWHM is consistent with the typical velocity width of other molecular lines except for CO and HCO$^+$. 

The derived column densities and upper limits (2 $\sigma$ level) are summarized in Table \ref{tab_N}. 

\begin{deluxetable}{ l c c c c c c c c c c}
\tablecaption{Line Parameters \label{tab_lines}}
\tablewidth{0pt}
\tabletypesize{\scriptsize} 
\tablehead{
\colhead{Molecule}   & \colhead{Transition}                               & \colhead{$Eu/k$} & \colhead{Frequency} & \colhead{$T_{mb}$} & \colhead{$\Delta$$V$} & \colhead{$\int T_{mb} dV$} & \colhead{$V_{LSR}$} & \colhead{RMS} & \colhead{$I_{\mathrm{time}}$\tablenotemark{a}} & \colhead{Note} \\
\colhead{ }          & \colhead{ }                                        & \colhead{(K)}    & \colhead{(GHz)}     & \colhead{(K)}      & \colhead{(km/s)}      & \colhead{(K km/s)}         & \colhead{(km/s)}    & \colhead{(K)} & \colhead{(s)}      & \colhead{}
}
\startdata
 CO                  &  3--2                                              & 33         &        345.7960 & $\sim$65             & $\sim$16        & $\sim$1150           & $\sim$252       & 0.42 & \ddag & 1  \\
 C$^{17}$O           &  3--2                                              & 32         &        337.0611 & 2.30 $\pm$ 0.08      & 5.6             & 13.77 $\pm$ 0.98     & 251.1           & 0.21 & \dag  &    \\
 HCO$^+$             &  4--3                                              & 43         &        356.7342 & 29.87 $\pm$ 0.08     & 8.5             & 270.86 $\pm$ 1.73    & 250.6           & 0.21 & \ddag &    \\
 H$^{13}$CO$^+$      &  4--3                                              & 42         &        346.9983 & 1.59 $\pm$ 0.06      & 7.3             & 12.34 $\pm$ 1.07     & 250.5           & 0.20 & \ddag &    \\
 NO                  &  7/2,9/2--5/2,7/2 $f$                              & 36         &        350.6895 & 1.63 $\pm$ 0.06      & 7.2             & 12.53 $\pm$ 0.97     & 250.6           & 0.19 & \dag  & 2  \\
                     &  7/2,7/2--5/2,5/2 $f$                              & 36         &        350.6908 &                      &                 &                      &                 &      &       &    \\
                     &  7/2,5/2--5/2,3/2 $f$                              & 36         &        350.6948 &                      &                 &                      &                 &      &       &    \\
 NO                  &  7/2,9/2--5/2,7/2 $e$                              & 36         &        351.0435 & 0.85 $\pm$ 0.08      & 3.8             & 3.44 $\pm$ 0.69      & 251.7           & 0.19 & \dag  &    \\
 NO                  &  7/2,7/2--5/2,5/2 $e$                              & 36         &        351.0515 & 0.71 $\pm$ 0.07      & 6.5             & 4.86 $\pm$ 0.97      & 251.7           & 0.19 & \dag  & 3  \\
                     &  7/2,5/2--5/2,3/2 $e$                              & 36         &        351.0517 &                      &                 &                      &                 &      &       &    \\
 H$_2$CO             &  5$_{1,5}$--4$_{1,4}$                              & 62         &        351.7686 & 2.48 $\pm$ 0.07      & 6.5             & 17.10 $\pm$ 1.02     & 250.7           & 0.20 & \dag  &    \\
 CH$_3$OH            &  7$_0$--6$_0$ A$^+$                                & 65         &        338.4087 & $<$0.30              & \nodata         & $<$1.9               & \nodata         & 0.15 & \dag  &    \\
 HNCO                &  16$_{0,16}$--15$_{0,15}$                          & 143        &        351.6333 & $<$0.28              & \nodata         & $<$1.8               & \nodata         & 0.14 & \dag  &    \\
 SiO                 &  8--7                                              & 75         &        347.3306 & 1.64 $\pm$ 0.06      & 7.1             & 12.32 $\pm$ 1.05     & 250.4           & 0.20 & \ddag &    \\
 C$^{34}$S           &  7--6                                              & 65         &        337.3965 & $<$0.30              & \nodata         & $<$1.9               & \nodata         & 0.15 & \dag  &    \\
 H$_2$CS             &  10$_{1,10}$--9$_{1,9}$                            & 102        &        338.0832 & 0.44 $\pm$ 0.07      & 3.2             & 1.49 $\pm$ 0.50      & 249.6           & 0.21 & \dag  &    \\
 CH$_3$OCH$_3$       &  8$_{4,5}$--7$_{3,4}$ AE                           & 55         &        356.5753 & $<$0.30              & \nodata         & $<$1.9               & \nodata         & 0.15 & \ddag & 3  \\
                     &  8$_{4,5}$--7$_{3,4}$ EE                           & 55         &        356.5760 &                      &                 &                      &                 &      &       &    \\
 C$_2$H$_5$OH        &  10$_{4,7}$--9$_{3,6}$                             & 66         &        357.0674 & $<$0.30              & \nodata         & $<$1.9               & \nodata         & 0.15 & \ddag &    \\
 $^{33}$SO           &  8$_7$--7$_6$                                      & 81         &        337.1986 & 2.84 $\pm$ 0.07      & 5.8             & 17.63 $\pm$ 1.00     & 250.9           & 0.21 & \dag  & 4  \\
 HC$_3$N             &  38--37                                            & 323        &        345.6090 & $<$0.27              & \nodata         & $<$1.7               & \nodata         & 0.13 & \ddag &    \\
 HCOOCH$_3$          &  9$_{9,0}$--8$_{8,1}$ A                            & 80         &        345.7187 & $<$0.27              & \nodata         & $<$1.7               & \nodata         & 0.13 & \ddag & 3  \\
                     &  9$_{9,1}$--8$_{8,0}$ A                            & 80         &        345.7187 &                      &                 &                      &                 &      &       &    \\
 SO$_2$              &  16$_{7,9}$--17$_{6,12}$                           & 245        &        336.6696 & 1.87 $\pm$ 0.07      & 7.8             & 15.50 $\pm$ 1.19     & 250.6           & 0.21 & \dag  &    \\
 SO$_2$              &  18$_{4,14}$--18$_{3,15}$                          & 197        &        338.3060 & 5.25 $\pm$ 0.07      & 6.2             & 34.54 $\pm$ 1.08     & 250.5           & 0.21 & \dag  &    \\
 SO$_2$              &  20$_{1,19}$--19$_{2,18}$                          & 199        &        338.6118 & 6.39 $\pm$ 0.07      & 6.7             & 45.67 $\pm$ 1.02     & 250.8           & 0.21 & \dag  &    \\
 SO$_2$              &  26$_{9,17}$--27$_{8,20}$                          & 521        &        345.4490 & 0.76 $\pm$ 0.07      & 3.4             & 2.76 $\pm$ 0.53      & 252.0           & 0.19 & \ddag &    \\
 SO$_2$              &  19$_{1,19}$--18$_{0,18}$                          & 168        &        346.6522 & 7.99 $\pm$ 0.07      & 6.5             & 54.93 $\pm$ 1.00     & 250.7           & 0.20 & \ddag &    \\
 SO$_2$              &  10$_{6,4}$--11$_{5,7}$                            & 139        &        350.8628 & 2.43 $\pm$ 0.06      & 5.8             & 15.10 $\pm$ 0.85     & 250.8           & 0.19 & \dag  &    \\
 SO$_2$              &  14$_{4,10}$--14$_{3,11}$                          & 136        &        351.8739 & 6.62 $\pm$ 0.07      & 6.6             & 46.24 $\pm$ 1.01     & 250.7           & 0.20 & \dag  &    \\
 SO$_2$              &  10$_{4,6}$--10$_{3,7}$                            & 90         &        356.7552 & 6.19 $\pm$ 0.07      & 7.0             & 46.21 $\pm$ 1.18     & 250.8           & 0.21 & \ddag &  5  \\
 SO$_2$              &  13$_{4,10}$--13$_{3,11}$                          & 123        &        357.1654 & 6.03 $\pm$ 0.07      & 6.0             & 38.25 $\pm$ 0.95     & 250.8           & 0.21 & \ddag &    \\
 SO$_2$ $\nu$$_2$=1  &  5$_{3,3}$--4$_{2,2}$                              & 781        &        357.0872 & 0.50 $\pm$ 0.07      & 3.0             & 1.56 $\pm$ 0.49      & 251.5           & 0.21 & \ddag &    \\
 $^{33}$SO$_2$       &  5$_{3,3}$--4$_{2,2}$                              & 35         &        346.5901 & 0.57 $\pm$ 0.06      & 6.2             & 3.76 $\pm$ 1.62      & 249.7           & 0.20 & \ddag & 6  \\
 $^{33}$SO$_2$       &  13$_{4,10}$--13$_{3,11}$                          & 122        &        350.7881 & 0.42 $\pm$ 0.05      & 4.1             & 1.83 $\pm$ 0.55      & 251.1           & 0.19 & \dag  & 7  \\
 $^{33}$SO$_2$       &  15$_{4,12}$--15$_{3,13}$                          & 149        &        350.9146 & 0.40 $\pm$ 0.06      & 4.5             & 1.89 $\pm$ 0.61      & 249.7           & 0.19 & \dag  & 7  \\
 $^{33}$SO$_2$       &  11$_{4,8}$--11$_{3,9}$                            & 99         &        350.9951 & 0.64 $\pm$ 0.05      & 5.3             & 3.64 $\pm$ 0.75      & 250.9           & 0.19 & \dag  & 7  \\
 $^{33}$SO$_2$       &  5$_{4,2}$--5$_{3,3}$                              & 52         &        351.6351 & 0.48 $\pm$ 0.06      & 4.2             & 2.16 $\pm$ 0.59      & 250.2           & 0.20 & \dag  & 7  \\
 $^{33}$SO$_2$       &  17$_{4,14}$--17$_{3,15}$                          & 179        &        351.7449 & $<$0.80              & \nodata         & $<$2.4               & \nodata         & 0.20 & \dag  & 7  \\
 $^{34}$SO$_2$       &  13$_{2,12}$--12$_{1,11}$                          & 92         &        338.3204 & 0.87 $\pm$ 0.06      & 5.6             & 5.14 $\pm$ 0.83      & 251.4           & 0.21 & \dag  &    \\
 $^{34}$SO$_2$       &  14$_{4,10}$--14$_{3,11}$                          & 134        &        338.7857 & 0.98 $\pm$ 0.06      & 6.9             & 7.26 $\pm$ 1.07      & 250.2           & 0.21 & \dag  &    \\
 $^{34}$SO$_2$       &  7$_{4,4}$--7$_{3,5}$                              & 64         &        345.5197 & 1.01 $\pm$ 0.07      & 6.1             & 6.59 $\pm$ 0.93      & 250.9           & 0.19 & \ddag &    \\
 $^{34}$SO$_2$       &  6$_{4,2}$--6$_{3,3}$                              & 57         &        345.5531 & 0.67 $\pm$ 0.06      & 5.6             & 4.03 $\pm$ 0.77      & 251.2           & 0.19 & \ddag &    \\
 $^{34}$SO$_2$       &  5$_{4,2}$--5$_{3,3}$                              & 52         &        345.6513 & 0.61 $\pm$ 0.05      & 5.6             & 3.66 $\pm$ 0.84      & 251.4           & 0.19 & \ddag &    \\
 $^{34}$SO$_2$       &  4$_{4,0}$--4$_{3,1}$                              & 47         &        345.6788 & 0.64 $\pm$ 0.08      & 3.7             & 2.49 $\pm$ 0.72      & 250.6           & 0.19 & \ddag &    \\
 $^{34}$SO$_2$       &  17$_{4,14}$--17$_{3,15}$                          & 179        &        345.9293 & 0.59 $\pm$ 0.05      & 5.0             & 3.16 $\pm$ 0.67      & 249.7           & 0.19 & \ddag &    \\
 $^{34}$SO$_2$       &  28$_{2,26}$--28$_{1,27}$                          & 391        &        347.4831 & 0.23 $\pm$ 0.06      & 3.3             & 0.78 $\pm$ 0.54      & 250.7           & 0.20 & \ddag &    \\
 $^{34}$SO$_2$       &  21$_{4,18}$--21$_{3,19}$                          & 250        &        352.0829 & 0.68 $\pm$ 0.07      & 3.6             & 2.62 $\pm$ 0.57      & 251.0           & 0.20 & \dag  &    \\
 $^{34}$SO$_2$       &  20$_{0,20}$--19$_{1,19}$                          & 185        &        357.1022 & 1.32 $\pm$ 0.07      & 5.2             & 7.27 $\pm$ 0.80      & 250.4           & 0.21 & \ddag &    \\
\enddata
\tablecomments{
\scriptsize{
Uncertainties and upper limits are of 2 $\sigma$ level and do not include systematic errors due to baseline subtraction and adopted spectroscopic constants. 
$^a$Total on-source integration time, where $\dag$ represents 575 seconds and $\ddag$ 1452 seconds. 
(1) Saturated. 
(2) Blended with three hyperfine components. 
(3) Blend with two hyperfine components. 
(4) Blend with four hyperfine components. 
(5) Partly blended with HCO$^+$(4--3), which is subtracted when fitting this SO$_2$ line. 
(6) Blend with seven hyperfine components. 
(7) Blend with ten hyperfine components. 
}}
\end{deluxetable}

\begin{deluxetable}{ l c c }
\tablecaption{Estimated column densities \label{tab_N}}
\tablewidth{0pt}
\tabletypesize{\footnotesize} 
\tablehead{
\colhead{Molecule}   & \colhead{$T$$_{rot}$}   &       \colhead{$N$(X)}            \\
\colhead{ }                & \colhead{(K)}                 &        \colhead{(cm$^{-2}$)} }
\startdata 
 SO$_2$                 &   190\tablenotemark{*}        &         8.4 $\pm$ 0.3 $\times$ 10$^{15}$\tablenotemark{*}     \\
 $^{34}$SO$_2$     &   95\tablenotemark{*}          &         6.2 $\pm$ 0.9 $\times$ 10$^{14}$\tablenotemark{*}   \\
 $^{33}$SO$_2$     &   64\tablenotemark{*}         &          2.1 $\pm$ 0.7 $\times$ 10$^{14}$\tablenotemark{*}     \\
\tableline
 H$_2$                     &   --           &         4.5 $\times$ 10$^{23}$    \\
 CO                          &   100        & $\sim$8 $\times$ 10$^{17}$     \\
 C$^{17}$O              &   100       &         1.2 $\pm$ 0.09 $\times$ 10$^{16}$     \\ 
 HCO$^+$                &   100       &        1.4 $\pm$ 0.01 $\times$ 10$^{14}$    \\ 
 H$^{13}$CO$^+$    &   100       &        6.9 $\pm$ 0.6 $\times$ 10$^{12}$      \\
 H$_2$CO               &   100       &        1.0 $\pm$ 0.1 $\times$ 10$^{14}$     \\
 CH$_3$OH            &   100       &   $<$3.5 $\times$ 10$^{14}$   \\
 NO                          &   100       &        9.1 $\pm$ 2.5 $\times$ 10$^{15}$    \\
 HNCO                    &   100       &   $<$4.3 $\times$ 10$^{13}$   \\ 
 HC$_3$N              &   100        &   $<$1.8 $\times$ 10$^{13}$   \\
 SiO                        &   100        &         1.5 $\pm$ 0.1 $\times$ 10$^{13}$    \\
 C$^{34}$S             &   100        &   $<$1.0 $\times$ 10$^{13}$   \\
 H$_2$CS               &   100        &         2.8 $\pm$ 0.9 $\times$ 10$^{13}$    \\
 $^{33}$SO             &   100        &         2.7 $\pm$ 0.6 $\times$ 10$^{14}$    \\
 CH$_3$OCH$_3$ &   100        &   $<$1.3 $\times$ 10$^{15}$   \\
 HCOOCH$_3$      &   100        &   $<$7.1 $\times$ 10$^{15}$   \\
 C$_2$H$_5$OH    &   100        &   $<$2.2 $\times$ 10$^{15}$   \\
\enddata
\tablecomments{
Uncertainties and upper limits are of 2 $\sigma$ level and do not include systematic errors due to adopted spectroscopic constants. \\
$^*$Derived based on the rotation diagram analysis (see $\S$ \ref{sec_rd} for details). 
}
\end{deluxetable}

\begin{deluxetable}{ l c c c c}
\tablecaption{Comparison of rotational temperatures  \label{tab_T}}
\tablecolumns{4}
\tablewidth{0pt}
\tabletypesize{\footnotesize} 
\tablehead{
\colhead{ }                      &        \multicolumn{4}{c}{$T$$_{rot}$ [K]}                                  \\
                                                                                            \cline{2-5}                                                                             \\
\colhead{Molecule}       & \colhead{ST11\tablenotemark{a}}       & \colhead{Orion-KL\tablenotemark{b}}         & \colhead{W3 (H$_2$O)\tablenotemark{c}}     &    \colhead{G34.3+0.15\tablenotemark{d}}
}
\startdata 
 SO$_2$                       &  190 $\pm$ 5                        &         124 (121 $\pm$ 4)\tablenotemark{*}   &       184                                 &        108, 284\tablenotemark{**}   \\
 $^{34}$SO$_2$           &  95 $\pm$ 7                          &         138 (114 $\pm$ 6)\tablenotemark{*}   &        179                                 &        131   \\
 $^{33}$SO$_2$           &  64 $\pm$ 14                         &        104 (73 $\pm$ 5)\tablenotemark{*}     &       \nodata                            &        \nodata          \\
\enddata
\tablecomments{
Uncertainties are of 2 $\sigma$ level. \\
$^*$Numbers in parentheses in the Orion data are derived in this work (see $\S$ \ref{sec_rd} for details). 
$^{**}$Two temperature components. 
}
\tablerefs{
$^a$This work; 
$^b$\citet{Schi97}; 
$^c$\citet{Hel97}; 
$^d$\citet{Mac96}
}
\end{deluxetable}

\subsection{Column density of H$_2$ and total gas mass} \label{sec_N_H2}
A column density of molecular hydrogen, which usually dominate the total mass of embedded sources, is estimated using the dust continuum emission data obtained in our observations. 
The flux density of dust continuum, $F_{\nu}$, at the optically thin frequency, $\nu$, is expressed as 
\begin{equation}
F_{\nu} = \Omega \tau_{\nu} B_{\nu}(T_{d}), \label{Eq4}
\end{equation}
where $\Omega$ is a beam solid angle, $\tau_{\nu}$ is an optical depth, $B_{\nu}(T_{d})$ is the Planck function and $T_{d}$ is a dust temperature \citep{Whi92}. 
The optical depth is expressed as 
\begin{equation}
\tau_{\nu} = \rho_{d} \kappa_{\nu} L, \label{Eq5}
\end{equation}
where $\rho_{d}$ is a mass density of dust, $\kappa_{\nu}$ is a mass absorption coefficient, and $L$ is a path length. 
We use the mass absorption coefficient of dust grains coated by thin ice mantles as presented in \citet{Oss94}. 
Using the dust-to-gas mass ratio, $Z$, the mass density of dust is expressed as 
\begin{equation}
\rho_{d} = Z \mu \rho_{\mathrm{H}} = Z \mu N_{\mathrm{H}} m_{\mathrm{H}} / L, \label{Eq6}
\end{equation}
where $\mu$ is a mean atomic mass per hydrogen, $\rho_{H}$ is a mass density of hydrogen, $N_{H}$ is a column density of hydrogen, and $m_{H}$ is a hydrogen mass. 
We here assume $\mu$ to be 1.41 according to \citet{Cox00}. 
We assume that the dust-to-gas mass ratio in the LMC is lower than the typical Galactic value of 0.008 by a factor of three according to \citet{Ber08}, and we use $Z$ = 0.0027 in this work. 
By combining Equations \ref{Eq4}--\ref{Eq6}, and assuming that all the hydrogen is in the form of H$_2$ ($N_{H_2}$ = $N_{H}/2$), the column density of molecular hydrogen is derived by the following equation: 
\begin{equation}
N_{\mathrm{H_2}} = \frac{F_{\nu} / \Omega}{2 \kappa_{\nu} B_{\nu}(T_{d}) Z \mu m_{\mathrm{H}}}. \label{Eq7}
\end{equation}

We measure the flux density per beam solid angle, $F_{\nu}/\Omega$, using the 840 $\mu$m dust continuum image of ST11 shown in Figure \ref{image_continuum}. 
With the aperture size of 0.5$\arcsec$ in diameter, the $F_{\nu}/\Omega$ is measured to be 0.34 Jy/beam. 
For dust temperature, we assume $T_{d}$ = 40 K, which is a typical dust temperature of high-mass YSOs in the LMC estimated based on far-infrared observations \citep{vanL10}. 
The assumed dust temperature is consistent with the far-infrared SED peak of ST11 (see $\S$ \ref{sec_LM} for description of the SED). 
Using Equation \ref{Eq7} and the above parameters, we estimate $N_{H_2}$ = 4.5 $\times$ 10$^{23}$ cm$^{-2}$. 
This H$_2$ column density corresponds to the total gas mass of 115 M$_{\sun}$ in the line of sight. 
The estimated dust opacity at 840 $\mu$m is $\tau_{840 \mu m}$ = 0.011, which corresponds to the visual extinction ($A\mathrm{v}$) of $\sim$150 mag. 

We also estimate a lower limit on the gas density around ST11. 
We assume that gas is spherically distributed around a protostar with a radius which is the same as the beam size. 
With this assumption, the total gas mass around ST11 derived above corresponds to the H$_2$ number density of 2 $\times$ 10$^6$ cm$^{-3}$. 
We emphasize that this is a lower limit because the H$_2$ density increases inversely with the assumed source size.

\subsection{Luminosity and stellar mass} \label{sec_LM}
\begin{figure*}[!]
\begin{center}
\includegraphics[width=15.0cm]{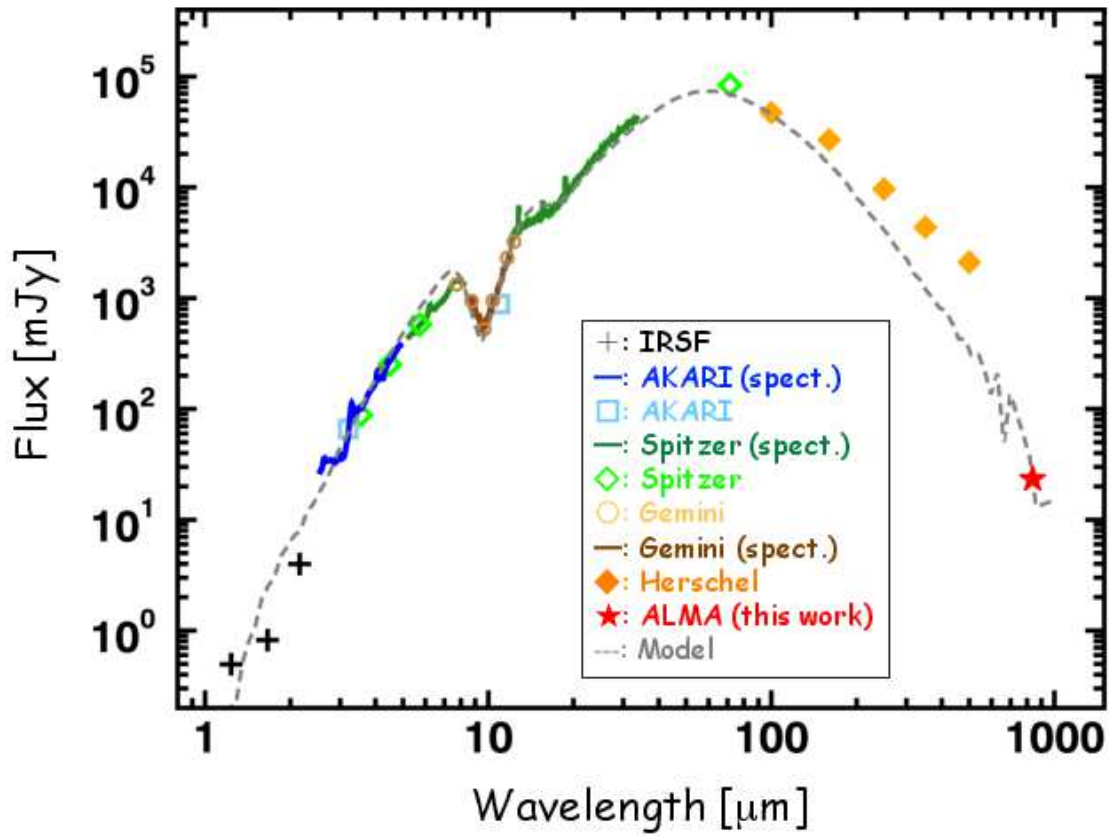}
\caption{
The spectral energy distribution of ST11. 
The plotted data are based on IRSF/SIRIUS photometry (pluses, black), {\it AKARI}/IRC spectroscopy (solid line, blue), {\it AKARI}/IRC photometry (open squares, light blue), {\it Spitzer}/MIPS spectroscopy (solid line, green), {\it Spitzer}/IRAC and MIPS photometry (open diamonds, light green), Gemini/T-ReCS narrow-band photometry (open circles, light brown), Gemini/T-ReCS N-band spectroscopy (solid line, brown), {\it Herschel}/PACS and SPIRE photometry (filled diamonds, orange), ALMA 840 $\mu$m continuum (filled star,red), and the best fitted SED model (dashed line, grey). 
See section $\S$ \ref{sec_LM} for details. 
}
\label{SED}
\end{center}
\end{figure*}

The SED of ST11 is shown in Figure \ref{SED}. 
Details of the collected data are summarized in Table \ref{tab_photo}. 
Most of the energy is emitted in the mid- to far-infrared wavelength regions and the peak of the SED is between 60 $\mu$m and 70 $\mu$m, which is consistent with the characteristics of high-mass YSOs. 

The bolometric luminosity of ST11 is estimated to be 5 $\times$ 10$^5$ L$_{\sun}$, which is derived by integrating the interpolated SED from 1 $\mu$m to 1000 $\mu$m. 
About 60 $\%$ of the total luminosity is emitted in in the 30--100 $\mu$m wavelength region. 
The stellar mass of ST11 is estimated using the Online SED Fitter\footnote{http://caravan.astro.wisc.edu/protostars/sedfitter.php} \citep{Rob07}. 
As the input data of the SED fit, we use 2--840 $\mu$m photometric and spectroscopic data, which are obtained from the IRSF/SIRIUS, {\it Spitzer} SAGE, {\it AKARI} LSLMC, and {\it Herschel} HERITAGE databases \citep{Kat07,Mei06,Kem10,Kat12,ST10,Mei13}. 
We also use the results of our mid-infrared narrow-band filter photometry (centered at 7.73, 8.74, 9,69, 10.38, 11.66, and 12.33 $\mu$m) and spectroscopy (8--12 $\mu$m) conducted with T-ReCS at the Gemini South telescope. 
The 840 $\mu$m flux is estimated using the present ALMA data (see $\S$ \ref{sec_N_H2}). 
We exclude the SPIRE 350 $\mu$m and 500 $\mu$m band data in the fit because they are possibly contaminated by diffuse emission around the YSO due to their large point spread function (about 27$\arcsec$ and 41$\arcsec$ in FWHM, respectively). 
The distance to ST11 is assumed to be the same as the distance to the LMC. 

The estimated stellar mass of ST11 in the best-fit model is 50 M$_{\sun}$. 
The visual extinction derived in the best-fit model is Av $\sim$80 mag, which differs by a factor of two from the value estimated based on the dust continuum data presented in $\S$ 4.4. 
The difference may arise from the assumption of optical properties of dust and/or the existence of a dust temperature gradient in the line of sight, though this does not affect the main conclusion of this paper. 
The result of the SED fit is plotted in Figure \ref{SED}. 

\begin{deluxetable}{ l c c c c}
\tablecaption{Photometric and spectroscopic data of ST11 \label{tab_photo}}
\tablewidth{0pt}
\tabletypesize{\footnotesize} 
\tablehead{
\colhead{Instrument}   & \colhead{Wavelength}   &  \colhead{Flux}     &  \colhead{FWHM\tablenotemark{a}}        & \colhead{Ref.} \\
\colhead{ }                   & \colhead{($\mu$m)}      &   \colhead{(mJy)}  &  \colhead{($\arcsec$)}  & 
}
\startdata 
IRSF/SIRIUS J                          &  1.23    &   0.50 $\pm$ 0.07   &  1.3   & 1  \\
IRSF/SIRIUS H                         &  1.66    &   0.82 $\pm$ 0.06   &   1.2   & 1  \\
IRSF/SIRIUS K$_\mathrm{s}$  &  2.16    &  3.98 $\pm$ 0.22   &    1.1   & 1  \\
\textit{AKARI}/IRC NG                             & 2.5-5    &   spectroscopy        &    7.3   & 2\tablenotemark{b}  \\
\textit{AKARI}/IRC N3                            & 3.2       & 66.0 $\pm$ 2.6        &    4.0   & 3  \\
\textit{Spitzer}/IRAC Band 1                    & 3.6      & 88.1 $\pm$ 6.6         &   1.7  & 4   \\
\textit{Spitzer}/IRAC Band 2                   & 4.5      & 252.2 $\pm$ 10.3     &    1.7   & 4  \\
\textit{Spitzer}/IRAC Band 3                   & 5.7      &  577.8 $\pm$ 23.6    &    1.9   & 4  \\
\textit{Spitzer}/IRAC Band 4                  & 7.9       &  1090.0 $\pm$ 39.7  &    2.0   & 4  \\
\textit{AKARI}/IRC S11                         & 11        & 885.6 $\pm$ 29.7     &    4.8   & 3  \\
Gemini-S/T-ReCS Si1                   & 7.7        & 1329 $\pm$ 29        &     0.6   & 5  \\
Gemini-S/T-ReCS Si2                   & 8.7        & 940 $\pm$ 53          &     0.5   & 5  \\
Gemini-S/T-ReCS Si3                   & 9.7        & 523 $\pm$ 10          &     0.6   & 5  \\
Gemini-S/T-ReCS Si4                   & 10.4      & 950 $\pm$ 10          &     0.8   & 5  \\
Gemini-S/T-ReCS Si5                   & 11.7      & 2297 $\pm$ 12         &    0.8   & 5  \\
Gemini-S/T-ReCS Si6                   & 12.3      & 3226 $\pm$ 28        &     0.5   & 5  \\
Gemini-S/T-ReCS Lo-Res             & 8--12     &  spectroscopy          &     0.7\tablenotemark{c}   & 5  \\
\textit{Spitzer}/IRS SL, SH, LH             & 5--36    &   spectroscopy           &    4-11\tablenotemark{c}   & 6  \\
\textit{Spitzer}/MIPS 70                        & 70         & 83810 $\pm$ 1313    &    18   & 4  \\
\textit{Herschel}/PACS 100                           & 100      & 47180 $\pm$ 3090    &      8.6   & 7  \\
\textit{Herschel}/PACS 160                          & 160      & 26800 $\pm$ 1629    &     13   & 7  \\
\textit{Herschel}/SPIRE 250                         & 250      & 9650 $\pm$ 589        &     18   & 7  \\
\textit{Herschel}/SPIRE 350                         & 350      & 4359 $\pm$ 226       &       27  & 7   \\
\textit{Herschel}/SPIRE 500                        & 500      & 2110 $\pm$ 151        &      41   & 7  \\
ALMA Band 7                            & 837      & 23.2 $\pm$ 2.32       &      0.5   & 5  
\enddata
\tablecomments{
$^a$FWHM of the point spread function. 
$^b$Extraction width of the slitless spectroscopic data. 
$^c$Slit width. 
}
\tablerefs{
(1) \citet{Kat07}; 
(2) \citet{ST10}; 
(3) \citet{Kat12}; 
(4) \citet{Mei06}; 
(5) This work; 
(6) \citet{Kem10}; 
(7) \citet{Mei13}
}
\end{deluxetable}

\section{Discussions} \label{sec_discussions} 
\subsection{Hot molecular core associated with ST11} \label{sec_hotcore} 
Hot cores are one of the early stages of high-mass star formation. 
It is suggested that the hot cores are in a transitional evolutionary stage between a deeply embedded protostellar object and a zero-age main-sequence star with a compact \ion{H}{2} region \citep[e.g.,][]{ZY07}. 
Physical properties of hot cores are characterized by a small source size ($\leq$0.1 pc), a high density ($\geq$10$^6$ cm$^{-3}$), and warm gas/dust temperature ($\geq$100 K) \citep[e.g.,][]{Kur00,vdT04}. 
In this section, we discuss the presence of a hot core around ST11 on the basis of its physical and spectral properties.

\subsubsection{Source size} \label{sec_size} 
A source size of ST11, which is measured by detected emission lines, is consistent with those of Galactic hot cores. 
The molecular lines of NO, SiO, $^{33}$SO, SO$_2$, $^{34}$SO$_2$, and $^{33}$SO$_2$ as well as dust continuum show a compact source size, which is equal to or lower than the beam size of $\sim$0.1 pc. 
The lines of CO, C$^{17}$O, HCO$^+$, H$^{13}$CO$^+$ and H$_2$CO show somewhat extended distributions, but the dominant emission come from the compact region associated with a high-mass YSO. 
The extended components of these emission lines may arise from their relatively low critical densities and/or efficient formation routes in gas-phase reactions. 
Protostellar outflows, which is discussed in $\S$ \ref{sec_outflow}, could also affect the spatial distribution of these molecular gas. 

The spatial association of the molecular emission lines to the infrared source is another key to the nature of detected warm and dense molecular gas. 
Separation between a hot core region and an infrared source is reported to be smaller than $\sim$0.03 pc for typical Galactic sources \citep{DeB03}. 
With the present spatial resolution, the distribution of molecular lines and dust continuum should coincide if they arise from a hot core region. 
This is clearly observed in the present data as shown in Figure \ref{image_Others}.

\subsubsection{Density} \label{sec_density} 
The H$_2$ gas density around ST11 is estimated to be $>$2 $\times$ 10$^6$ cm$^{-3}$ in $\S$ \ref{sec_N_H2}. 
The value is a lower limit and actually clear detections of molecular lines with high critical densities such as those of SO$_2$ \citep[$n_{cr}$ $\sim$10$^7$ cm$^{-3}$,][]{Wil13} imply an even higher gas density around ST11. 
Galactic hot cores typically show gas densities between 10$^6$--10$^8$ cm$^{-3}$ \citep[estimated from Tab.1 presented in][]{Kur00}. 
The gas density around ST11 is thus consistent with those of known hot cores. 

\subsubsection{Gas temperature} \label{sec_temperature} 
Rotational temperatures of molecular gas around ST11 are compared with those of Galactic hot cores in Table \ref{tab_T}. 
The temperature of SO$_2$, $T_{rot}$ = 195 $\pm$ 5 K, is in good agreement with temperatures of three Galactic hot cores in the table, which are between 108 K and 284 K. 
The temperature of $^{34}$SO$_2$, $T_{rot}$ = 95 $\pm$ 7 K, is somewhat lower than that of SO$_2$, but this may be attributed to the combined effect of the lack of available higher excitation lines of $^{34}$SO$_2$ and slightly larger optical thickness of SO$_2$ lines. 
Note that the temperature of $^{33}$SO$_2$ is even lower, but this is probably because the detected lines are biased to those with low excitation temperatures ($E_u$ $<$ 200 K) due to the limited frequency coverage of the data. 
The rotation analysis of $^{34}$SO$_2$ and $^{33}$SO$_2$ for the Orion data using the similar set of lines with those used in the analysis of ST11 results in $T_{rot}$ = 114 $\pm$ 6 K for $^{34}$SO$_2$ and $T_{rot}$ = 73 $\pm$ 5 K for $^{33}$SO$_2$, which are lower than the values derived by \citet{Schi97} using a larger number of lines with a wide range of upper state energies (Tab. \ref{tab_T}). 
Therefore we conclude that the actual temperature of SO$_2$ gas is in the order of $\sim$100 K, which can trigger hot core chemistry via sublimation of ice mantles. 
Sublimation of ice mantles around ST11 is also suggested from the infrared observations of ices in \citet{ST10}, which reported that ST11 shows the second weakest ice absorption bands among 12 high-mass YSOs in the LMC.

\subsubsection{Central Protostar} \label{sec_protostar} 
As estimated in $\S$ \ref{sec_LM}, the bolometric luminosity (5 $\times$ 10$^5$ L$_{\sun}$) and stellar mass (50 M$_{\sun}$) of ST11 are consistent with the properties of high-mass YSOs. 
This indicates that ST11 can form a hot core region with the help of its intense radiation. 
In addition, the red SED of ST11 (Fig. \ref{SED}) indicates that it is still in an early evolutionary stage, which is consistent with the properties of hot core sources.

\subsubsection{Spectral properties} \label{sec_spectal} 
Velocity widths of emission lines from ST11 are typically 4--7 km/s except for CO and HCO$^+$. 
This is in good agreement with velocity widths of emission lines from Galactic hot core regions \citep[typically 4--10 km/s, e.g.,][]{Hel97}. 
Systemic velocities of the lines are in a very narrow range, typically 250--251 km/s, which suggests that the detected molecular species spatially coexist in a small region around a high-mass YSO. 

SO$_2$ is often detected in hot core sources and one of the useful tracers of warm and dense gas around a high-mass YSO \citep[e.g.,][]{Beu09}. 
However, sometimes warm SO$_2$ gas is also prominently detected in deeply embedded and even younger protostellar objects in which hot cores are not yet formed \citep[e.g., W3 IRS5 in][]{Hel94}. 
The reason of such SO$_2$ enhancement is suggested to be due to shock dominated chemistry triggered by protostellar outflows. 
These deeply embedded sources often show a deep self-absorption profile in their CO or HCO$^+$ emission lines due to the presence of a significant amount of cold gas in the envelope. 
As shown in Figure \ref{line_Others}, emission from ST11 does not show such a deep self-absorption profile. 
This suggests that emission from ST11 is mostly dominated by warm gas and that ST11 has already reached the hot core phase. 

On the other hand, relatively weak strengths of hydrogen recombination lines seen in the near-infrared spectrum of ST11 suggests that a prominent \ion{H}{2} region is not yet formed around the source \citep[see Fig.1 in][]{ST10}. 
Such a transitional evolutionary phase is well consistent with the properties of hot cores. 
\\

In summary, the compact size of the emitting source, the presence of warm and dense molecular gas around a high-mass protostar, the sublimation of ice mantles, and the detections of rich molecular lines suggest that ST11 is associated with a hot molecular core. 
This is the first detection of an extragalactic hot molecular core. 
Note that CH$_3$OH and complex organic molecules, which are often detected in Galactic hot cores, are not detected in ST11. 
The reason for the lack of these molecular species is discussed in the next section in the context of characteristic hot core chemistry in low metallicity.

\subsection{Molecular abundances} \label{sec_X_mol} 
Hot cores play a key role in the chemical complexity of interstellar and circumstellar molecules. 
In this section, we compare the chemical compositions of ST11 with those of Galactic hot cores and discuss the impact of metallicity on chemical processes in hot cores. 

Fractional abundances of molecules around ST11 are shown in Table \ref{tab_X}. 
We use the following isotope abundances for ST11; $^{12}$C/$^{13}$C = 49, $^{16}$O/$^{17}$O = 3400, and $^{32}$S/$^{34}$S = 15, according to \citet{Wan09} in which isotope ratios for the star-forming region N113 in the LMC are reported. 
For $^{33}$S, we use $^{32}$S/$^{33}$S = 40, which is estimated in this work (see $\S$ \ref{Sec_SO2}). 
For a comparison purpose, the table also includes the abundances of three Galactic hot cores, Orion, W3 (H$_2$O), and G34.3+0.15. 
Their abundances are re-estimated using the following isotope ratios of the local ISM and the Sun: $^{12}$C/$^{13}$C = 77, $^{16}$O/$^{17}$O = 1800, $^{32}$S/$^{34}$S = 22, and $^{32}$S/$^{33}$S = 127, according to \citet{Wil94} and \citet{And89}. 
All the abundances are estimated from the 345 GHz region data, except for NO in Orion, which is estimated using the transitions near 150 GHz. 

Since ST11 is observed with the beam size of 0.12 pc, we here compare the abundances which are derived from the data with similar beam sizes. 
For Orion, we calculate the average abundance of five regions around the hot core (hot core, compact ridge, extended ridge, northwest plateau, and southeast plateau) using the data presented in \citet{Sut95}. 
The abundances of NO and HNCO in Orion are only for the central part of the hot core and the data are adopted from \citet{Ziu91} and \citet{Schi97}, respectively. 
The averaged Orion abundances roughly reproduce the abundances observed for the $\sim$0.08 pc region in diameter. 
The beam sizes of W3 (H$_2$O) and G34.3+0.15 correspond to 0.13 pc and 0.21 pc, respectively. 
We here assume the distances to Orion, W3 (H$_2$O), and G34.3+0.15 to be 0.41 kpc, 1.95 kpc, and 3.1 kpc \citep{Mac96,Rei09}. 

\begin{figure*}[!t]
\begin{center}
\includegraphics[width=16.3cm]{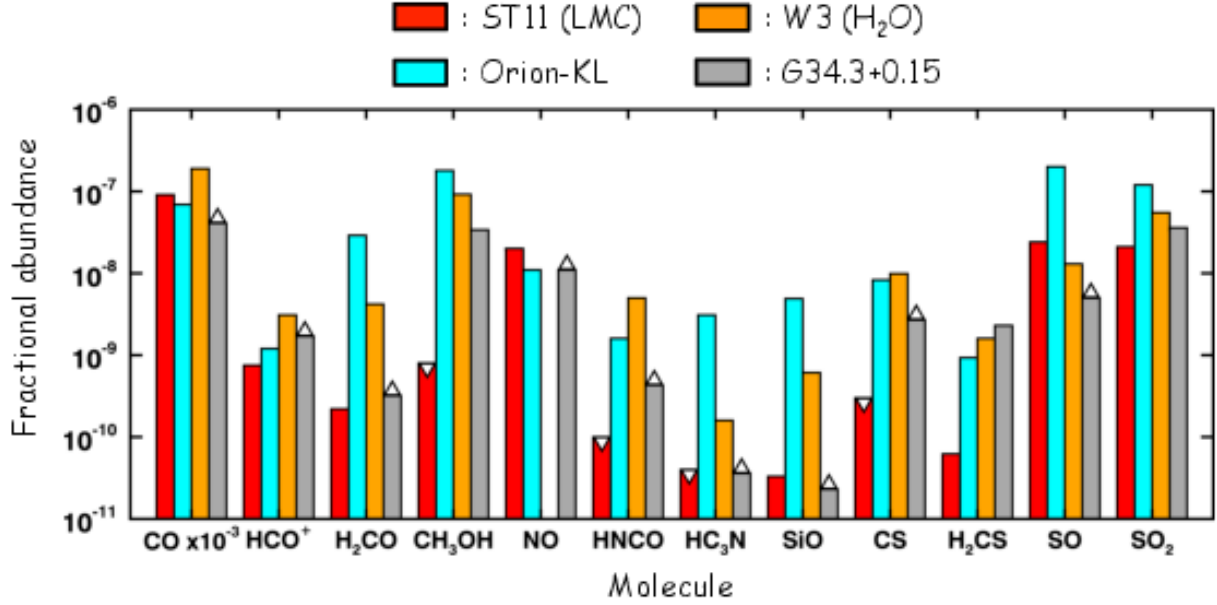}
\caption{
Molecular abundances of CO, HCO$^+$, H$_2$CO, CH$_3$OH, NO, HNCO, HC$_3$N, SiO, CS, H$_2$CS, SO, and SO$_2$ for ST11 in the LMC (red) and Galactic hot cores, Orion hot core (cyan), W3 H$_2$O (orange), and G34.3+0.15 (grey). 
The downward and upward triangles represent the upper and lower limits, respectively. 
The plotted data are summarized in Table \ref{tab_X}. 
}
\label{histo_ab}
\end{center}
\end{figure*}

Figure \ref{histo_ab} compares fractional abundances of ST11 and Galactic hot cores. 
In general, most of the molecular species in ST11 show lower abundances compared to Galactic hot cores. 
Several molecules such as H$_2$CO, CH$_3$OH, HNCO, and CS show significantly lower abundances, which can not be simply explained by the low abundances of heavy elements in the LMC. 
On the other hand, NO shows the higher abundance in ST11 than in Galactic sources despite the notably low nitrogen abundance in the LMC. 
Characteristics of individual molecules are discussed below.

\begin{deluxetable}{ l c c c c c}
\tablecaption{Fractional abundances  \label{tab_X}}
\tablecolumns{5}
\tablewidth{0pt}
\tabletypesize{\small} 
\tablehead{
\colhead{ }                      &        \multicolumn{4}{c}{$N$(X)/$N$(H$_2$)}                                                                                                                                           & \colhead{ }    \\
                                                                                            \cline{2-5}                                                                             \\
\colhead{Molecule}     & \colhead{ST11\tablenotemark{a}} & \colhead{Orion\tablenotemark{b}} & \colhead{W3 (H$_2$O)\tablenotemark{c}} & \colhead{G34.3+0.15\tablenotemark{d}} & \colhead{Note}
}
\startdata 
 CO                                &   9.1 $\pm$ 0.7 $\times$ 10$^{-5}$            &        6.9 $\times$ 10$^{-5}$     &        1.9 $\times$ 10$^{-4}$     &       $>$4.1 $\times$ 10$^{-5}$      &  1   \\
 HCO$^+$                      &   7.5 $\pm$ 0.7 $\times$ 10$^{-10}$         &        1.2 $\times$ 10$^{-9}$     &        3.1 $\times$ 10$^{-9}$     &       $>$1.7 $\times$ 10$^{-9}$      &  2   \\
 H$_2$CO                     &   2.2 $\pm$ 0.2 $\times$ 10$^{-10}$          &        2.9 $\times$ 10$^{-8}$     &        4.2 $\times$ 10$^{-9}$      &       $>$3.2 $\times$ 10$^{-10}$    &     \\
 CH$_3$OH                  &   $<$8 $\times$ 10$^{-10}$                        &       1.8 $\times$ 10$^{-7}$     &        9.2 $\times$ 10$^{-8}$      &       3.4 $\times$ 10$^{-8}$              &      \\
 NO                                &   2.0 $\pm$ 0.6 $\times$ 10$^{-8}$           &         1.1 $\times$ 10$^{-8}$     &        \nodata                             &        $>$1.1 $\times$ 10$^{-8}$      &      \\
 HNCO                          &   $<$1 $\times$ 10$^{-10}$                        &          1.6 $\times$ 10$^{-9}$     &        5.0 $\times$ 10$^{-9}$      &       $>$4.3 $\times$ 10$^{-10}$     &     \\ 
 HC$_3$N                    &   $<$4 $\times$ 10$^{-11}$                         &         3.1 $\times$ 10$^{-9}$      &        1.6 $\times$ 10$^{-10}$    &      $>$3.6 $\times$ 10$^{-11}$       &      \\
 SiO                              &   3.3 $\pm$ 0.2 $\times$ 10$^{-11}$           &         4.9 $\times$ 10$^{-9}$     &        6.1 $\times$ 10$^{-10}$     &       $>$2.3 $\times$ 10$^{-11}$     &       \\
 CS                               &   $<$3 $\times$ 10$^{-10}$                        &         8.3 $\times$ 10$^{-9}$     &        9.9 $\times$ 10$^{-9}$      &       $>$2.7 $\times$ 10$^{-9}$       &  3    \\
 H$_2$CS                     &   6.2 $\pm$ 2.0 $\times$ 10$^{-11}$          &         9.3 $\times$ 10$^{-10}$      &        1.6 $\times$ 10$^{-9}$    &        2.3 $\times$ 10$^{-9}$            &       \\
 SO                               &   2.4 $\pm$ 0.8 $\times$ 10$^{-8}$            &         2.0 $\times$ 10$^{-7}$     &        1.3 $\times$ 10$^{-8}$     &        $>$5.0 $\times$ 10$^{-9}$      &  4     \\
 SO$_2$                       &   2.1 $\pm$ 0.3 $\times$ 10$^{-8}$            &         1.2 $\times$ 10$^{-7}$      &        5.5 $\times$ 10$^{-8}$    &        3.6 $\times$ 10$^{-8}$            &  5    \\
 CH$_3$OCH$_3$      & $<$3$\times$ 10$^{-9}$                               &         9.1 $\times$ 10$^{-9}$      &        2.1 $\times$ 10$^{-8}$    &        \nodata                           &         \\
 HCOOCH$_3$           & $<$2 $\times$ 10$^{-8}$                              &         1.6 $\times$ 10$^{-8}$      &        7.0 $\times$ 10$^{-9}$    &        3.0 $\times$ 10$^{-8}$  &         \\
 C$_2$H$_5$OH        & $<$5 $\times$ 10$^{-9}$                              &         8.6 $\times$ 10$^{-10}$      &      \nodata                            &        6.6 $\times$ 10$^{-9}$  &         \\
\enddata
\tablecomments{
Uncertainties and upper limits are of 2 $\sigma$ level and do not include systematic errors due to adopted spectroscopic constants. 
See $\S$ \ref{sec_X_mol} for details of the data. 
(1) Estimated from C$^{17}$O. 
(2) Estimated from H$^{13}$CO$^+$ except for Orion, which is estimated from HC$^{18}$O$^+$. 
(3) Estimated from C$^{34}$S. 
(4) Estimated from $^{34}$SO, except for ST11, which is estimated from $^{33}$SO. 
(5) Estimated from $^{34}$SO$_2$. 
}
\tablerefs{
$^a$This work; 
$^b$\citet{Ziu91,Sut95,Schi97}, also see $\S$ \ref{sec_X_mol}; 
$^c$\citet{Hel97}; 
$^d$\citet{Mac96}
}
\end{deluxetable}

\subsubsection{CO} \label{sec_CO} 
Carbon monoxide is detected by the CO(3--2) and C$^{17}$O(3--2) lines in ST11. 
The line intensity ratio of CO(3--2)/C$^{17}$O(3--2) is about 28, and using $^{16}$O/$^{17}$O = 3400 in the LMC \citep{Wan09} we estimate optical depths of $\sim$120 and $\sim$0.04 for CO(3--2) and C$^{17}$O(3--2), respectively. 
The CO(3--2) line is completely optically thick, while the C$^{17}$O(3--2) line is optically thin as in most of Galactic hot cores \citep[e.g.,][]{Hel97}. 
The abundance of CO in ST11 estimated from C$^{17}$O is 9.1 $\times$ 10$^{-5}$, which is very similar to the average CO abundance of 1.0 $\times$ 10$^{-4}$ for three Galactic hot cores in Table \ref{tab_X}. 
Elemental carbon and oxygen are about three times less abundant in the LMC; [C/H]$_{\mathrm{LMC}}$ = 1.2 $\times$ 10$^{-4}$ and [O/H]$_{\mathrm{LMC}}$ = 2.3 $\times$ 10$^{-4}$ \citep{Kor02}, while [C/H]$_{\sun}$ = 3.3 $\times$ 10$^{-4}$ and [O/H]$_{\sun}$ = 6.8 $\times$ 10$^{-4}$ \citep{Gre98}. 
Given the low abundances of carbon and oxygen in the LMC, the CO around ST11 is slightly overproduced as compared with Galactic hot cores.

\subsubsection{HCO$^+$} \label{sec_HCO} 
Formyl ion is detected via the HCO$^+$(4--3) and H$^{13}$CO$^+$(4--3) lines. 
The line intensity ratio of HCO$^+$(4--3)/H$^{13}$CO$^+$(4--3) is about 19, and using $^{12}$C/$^{13}$C = 49 in the LMC \citep{Wan09} we estimate optical depths of $\sim$2 and $\sim$0.05 for HCO$^+$(4--3) and H$^{13}$CO$^+$(4--3), respectively. 
The HCO$^+$(4--3) line is moderately optically thick, while the H$^{13}$CO$^+$(4--3) line is optically thin. 
The abundance of HCO$^+$ in ST11 estimated from H$^{13}$CO$^+$ is 7.5 $\times$ 10$^{-10}$, which is lower by a factor of $\sim$3 compared to the average HCO$^+$ abundance of 2.0 $\times$ 10$^{-9}$ for three Galactic hot cores in Table \ref{tab_X}. 

We here discuss the connection between the observed HCO$^+$ abundance and the cosmic-ray ionization rate in the LMC. 
One of the possible pathways to form HCO$^+$ in molecular clouds is the gas-phase reaction: 
\begin{equation} 
\mathrm{
H_{3}^{+} + CO \to HCO^{+} + H_{2}, 
}
\end{equation} 
in which H$_{3}^{+}$ is formed by ionization of H$_2$ via cosmic rays and subsequent reaction with H$_2$ \citep[e.g.,][]{Cas98}. 

We here simply assume that HCO$^+$ is a major cation and electron is a major anion in dense clouds. 
Given that production of cations and electrons by the cosmic-ray ionization balances with their recombination, the equilibrium is formulated as 
\begin{equation} 
\zeta n_{\mathrm{H_2}} = k_{rec} n_{\mathrm{HCO^+}} n_{e}, 
\end{equation} 
where $\zeta$ is a cosmic-ray ionization rate, $k_{rec}$ = 2.4 $\times$ 10$^{-7}$ (300/$T$)$^{0.69}$ cm$^3$ s$^{-1}$ (5.1 $\times$ 10$^{-7}$ at $T$ = 100 K) is a dissociative recombination rate of HCO$^+$ and electron \citep{Mit90}, $n$ indicates a number density of each species. 
With the assumption that cations and electrons have the the same abundance in neutral clouds, this equation is reformatted as
\begin{equation} 
X_{HCO^+} = X_e = \sqrt{\zeta/n_{H_2}k_{rec}}, \label{Eq_X_HCO}
\end{equation}
where $X$ indicates a fractional abundance relative to H$_2$. 
This equation indicates that the abundance of HCO$^+$ is proportional to $\zeta^{0.5}$ under the above simple assumption. 

The cosmic-ray density in the LMC is reported to be 20--30$\%$ of the typical Galactic value of $\zeta$ = 3 $\times$ 10$^{-17}$ s$^{-1}$ based on gamma-ray observations \citep{Abd10}. 
The low cosmic-ray density leads to the low cosmic-ray ionization rate \citep[e.g.,][]{Spi68}. 
According to Eq. \ref{Eq_X_HCO}, the HCO$^+$ abundance in the LMC is expected to be lower than the Galactic value by a factor of $\sim$2 for the same H$_2$ density and gas temperature. 
This factor is close to the observed difference of the HCO$^+$ abundance between ST11 and Galactic sources. 
Therefore, we speculate that the low cosmic-ray ionization rate in the LMC is one of the factors that contribute to the slightly low abundance of HCO$^+$ in ST11. 

Note that the above assumption is very simple; i.e., the cosmic-ray ionization rate may vary within the LMC and the electron abundance may differ from that of HCO$^+$ due to the presence of other ions. 
Furthermore, in the above discussion, we consider only one of the possible mechanisms to form HCO$^+$. 
In a circumstellar environment of hot cores, however, outflows and stellar radiation would affect the ionization condition. 
Previous studies actually argue that molecular outflows and shocks contribute to the enhancement of HCO$^+$ in star-forming cores \citep[e.g.,][]{Gir99,Raw04,Arc06}. 
Future observations of other molecular ions and shock tracers are highly required for comprehensive understanding of ionization conditions in the LMC.

\subsubsection{H$_2$CO} \label{sec_H2CO} 
Formaldehyde (H$_2$CO) is detected by the 5$_{1,5}$--4$_{1,4}$ transition at 351.7686 GHz ($E_u$ = 62 K). 
The abundance of H$_2$CO in ST11 is estimated to be 2.2 $\times$ 10$^{-10}$, while Galactic hot cores show the abundances between 10$^{-9}$ to 10$^{-8}$ as in Table \ref{tab_X}, except for G34.3+0.15 which only has a lower limit. 
The H$_2$CO is less abundant in ST11 than Galactic sources by 1--2 orders of magnitude. 

Both gas-phase and grain surface reaction pathways are suggested for the formation of H$_2$CO in dense ISM. 
The grain surface pathway for the H$_2$CO formation requires the hydrogenation of CO \citep[e.g.,][and references therein]{HW13}. 
However, this pathways is suppressed in the LMC due to higher dust temperatures (see detailed discussion presented in $\S$ \ref{sec_CH3OH}). 
Therefore we speculate that suppressed formation of the H$_2$CO ice on grain surfaces contributes to the low abundance of H$_2$CO gas in ST11.

\subsubsection{CH$_3$OH} \label{sec_CH3OH} 
Any of methanol (CH$_3$OH) lines is not detected in ST11. 
We estimate an upper limit on the fractional abundance to be $<$8 $\times$ 10$^{-10}$ using the spectrum at the frequency of the CH$_3$OH(7$_0$--6$_0$ A$^+$) transition at 338.4087 GHz ($E_u$ = 65 K). 
Abundances of CH$_3$OH in Galactic hot cores are typically between $\sim$10$^{-8}$ to $\sim$10$^{-7}$ as in Table \ref{tab_X}, but sometimes the abundance reaches $\sim$10$^{-6}$ \citep[e.g., G5.89-0.39 in ][]{Tho99}. 
In ST11, CH$_3$OH is depleted by at least 2--3 orders of magnitude as compared with Galactic hot cores. 

The low abundance of CH$_3$OH in the LMC has been suggested in previous studies. 
\citet{Nis15} reported, based on spectral line surveys toward molecular clouds in the LMC, that thermal emission lines of CH$_3$OH gas are significantly weak in the LMC compared to our Galaxy. 
Searches for maser emission in the LMC reported the underabundance of CH$_3$OH masers in the LMC \citep[e.g.,][]{Ell10}. 
Furthermore, \citet{ST16} reported that the CH$_3$OH ice is less abundant in the LMC than in our Galaxy based on infrared observations. 
They reported that all of the observed ten high-mass YSOs in the LMC show CH$_3$OH ice abundances less than 5--8 $\%$ relative to water ice, while about one-third of Galactic high-mass YSOs show CH$_3$OH ice abundances between 10 $\%$ and 40 $\%$. 
The authors suggested that warm ice chemistry is responsible for the low abundance of solid CH$_3$OH in the LMC; i.e., high dust temperatures in the LMC suppress the hydrogenation of CO on the grain surface, which leads to inefficient production of the CH$_3$OH ice. 
A decrease in the efficiency of CO hydrogenation at an elevated temperature is measured by laboratory experiments \citep[e.g.,][]{Wat03}. 
In addition, the reduced formation of CH$_3$OH in relatively warm molecular clouds is confirmed by numerical simulations of grain surface chemistry dedicated to the LMC environment \citep{Ach15}. 

The warm gas temperature around ST11 suggests that ice mantles are mostly sublimated. 
The low elemental abundances of carbon and oxygen in the LMC should partly contribute to the observed low CH$_3$OH abundance. 
However, we need additional explanation to account for the CH$_3$OH deficiency by several orders of magnitude, and this should be related to grain surface chemistry by which CH$_3$OH is mainly formed. 
We suggest that suppressed production of CH$_3$OH ice due to warm ice chemistry at the molecular cloud stage or the deeply embedded YSO stage is responsible for the deficiency of CH$_3$OH gas in this LMC hot core. 

The above interpretation on the low CH$_3$OH abundance can also be applied to the low abundance of H$_2$CO in ST11 since a major grain surface pathway of both species requires the CO hydrogenation. 
However, the difference is that H$_2$CO has possible formation routes both in the gas-phase and in the solid-phase, while CH$_3$OH does not have an efficient formation route in the gas-phase under typical molecular cloud conditions. 
We speculate that the H$_2$CO observed in ST11 is mainly produced by gas-phase reactions. 
A slightly extended distribution of the H$_2$CO emission (Fig. \ref{image_Others}) supports this idea, because the species sublimated from ice mantles often show the compact distribution compared to those produced in the gas phase. 
Owing to the lack of a possible gas-phase formation route, the degree of depletion around ST11 is actually larger in CH$_3$OH as compared to H$_2$CO.

\subsubsection{NO} \label{sec_NO} 
Nitric oxide (NO) is one of the interesting molecules that show a peculiar abundance in ST11. 
The abundance of NO in ST11 is estimated to be 2.0 $\times$ 10$^{-8}$. 
On the other hand, the average abundance and standard deviation of NO in six Galactic sources in \citet{Ziu91} including hot cores, high-mass protostellar objects, and Galactic center objects is 8.2 $\pm$ 2.9 $\times$ 10$^{-9}$. 
In the LMC, nitrogen is even more depleted than other major elements such as carbon and oxygen; the elemental abundance of nitrogen in the LMC is [N/H]$_{\mathrm{LMC}}$ = 1.0 $\times$ 10$^{-5}$ \citep{Kor02}, while [N/H]$_{\sun}$ = 8.3 $\times$ 10$^{-5}$ for the Sun \citep{Gre98}. 
Despite the lower nitrogen abundance by a factor of 8 in the LMC, the abundance of NO in ST11 is higher than Galactic star-forming regions by a factor of 2--3. 

Ammonia (NH$_3$) gas, which is often the most abundant nitrogen-bearing species in star-forming regions, is reported to be deficient in the LMC by 1.5--2 orders of magnitude as compared with Galactic star-forming regions \citep{Ott10}. 
A previous infrared study suggests that the NH$_3$ ice around an embedded high-mass YSO in the LMC is possibly less abundant as compared with Galactic high-mass YSOs \citep{ST16}. 
Although the NH$_3$ abundance in ST11 is unknown, the relatively high abundance of NO implies that NO could play an important role in nitrogen chemistry around ST11. 

It is suggested that NO is formed by a neutral-neutral reaction in the gas-phase \citep[e.g.,][]{Her73,Pin90}: 
\begin{equation} 
\mathrm{
N + OH \to NO + H, \label{Eq_NO_1}
}
\end{equation}
and destroyed by 
\begin{equation} 
\mathrm{
NO + N \to N_2 + O. \label{Eq_NO_2}
}
\end{equation}
A numerical simulation of nitrogen chemistry in warm molecular gas suggests that the resultant abundance of NO through the above reactions increases as the temperature of molecular gas increases \citep{Pin90}. 
However, the enhancement of the NO abundance is suggested to be only a factor of 1.5 as the gas temperature increases from 70 K to 250 K. 
Thus we speculate that pure gas-phase chemistry make only a limited contribution on the enhanced abundance of NO in ST11. 

A compact distribution of the NO emission observed in ST11 may hint at the possible origin in ice sublimation. 
Laboratory experiments argue that bombardment of energetic ions to interstellar ice analogues containing H$_2$O, O$_2$, and N$_2$ or CO and N$_2$ produce NO in ice mixtures \citep{Bod12,Sic12}. 
However, the low cosmic-ray density in the LMC (see $\S$ \ref{sec_HCO}) implies that the formation of NO by such energetic processing is less efficient around ST11. 
A reaction pathway though diffusive grain surface chemistry, N + O $\to$ NO, is also suggested in the literature \citep{TH82}, but uncertainty remain regarding the efficiency of the reaction in actual grain surfaces. 

The reason of the enhanced abundance of NO in ST11 despite the low elemental abundance of nitrogen in the LMC remains to be investigated. 
Detailed modeling of hot core chemistry in metal-poor environments is obviously necessary to interpret the enhanced abundance of NO in ST11.

\subsubsection{HNCO} \label{sec_HNCO} 
Isocyanic acid (HNCO) is not detected in ST11. 
We estimate an upper limit on the fractional abundance of HNCO to be $<$1 $\times$ 10$^{-10}$ based on the non-detection of the 16$_{0,16}$--15$_{0,15}$ transition at 351.6333 GHz ($E_u$ = 143 K). 
There seems to be an emission-like feature at the position of the HNCO(16$_{3,14}$--15$_{3,13}$ and 16$_{3,13}$--15$_{3,12}$, 351.4168 GHz) line with $V_{LSR}$ = 250.8 km/s. 
However, the high upper state energy of this transition ($E_u$ = 518 K) and the non-detection of the 16$_{0,16}$--15$_{0,15}$ transition suggest that this line is likely a spurious signal. 
Galactic hot cores and high-mass protostellar objects typically show the HNCO abundances between 10$^{-9}$ to 10$^{-8}$ \citep{Bis07}, which is consistent with the average abundance of 2.3 $\times$ 10$^{-9}$ for three Galactic hot cores in Table \ref{tab_X}. 
Thus the HNCO abundance in ST11 is lower than Galactic counterparts by at least 1--2 orders of magnitude. 

Both gas-phase and grain surface reaction pathways are suggested for the formation of HNCO in dense ISM \citep[e.g.,][]{TH82,Tur99}. 
The grain surface pathway requires the hydrogenation of CO to form HCO and subsequent reaction of HCO + N $\to$ HNCO. 
However, as discussed in $\S$ \ref{sec_CH3OH}, the hydrogenation of CO is suggested to be less efficient in the LMC due to high dust temperatures, which possibly suppress the grain surface formation of HNCO. 
This may contribute to the low abundance of HNCO in ST11 along with the low elemental abundance of nitrogen in the LMC.

\subsubsection{HC$_3$N}\label{sec_HC3N} 
Cyanoacetylene (HC$_3$N), an unsaturated carbon-chain molecule often detected in star-forming regions, is not detected in ST11. 
We estimate an upper limit on the fractional abundance of HC$_3$N to be $<$4 $\times$ 10$^{-11}$ based on the non-detection of the 38--37 transition at 345.6090 GHz ($E_u$ = 323 K). 
Abundances of HC$_3$N in Galactic hot cores vary from $\sim$10$^{-11}$ to $\sim$10$^{-9}$ as in Table \ref{tab_X}. 
The estimated upper limit on the HC$_3$N abundance in ST11 is thus at the lower end of the Galactic HC$_3$N abundance.

\subsubsection{SiO} \label{sec_SiO} 
Silicon monoxide (SiO) is detected by the J = 8--7 transition at 347.3306 GHz ($E_u$ = 75 K). 
The abundance of SiO for ST11 is 3.3 $\times$ 10$^{-11}$, which seems to be low compared to Galactic sources, but the dispersion in the Galactic SiO abundances is significantly large as shown in Figure \ref{histo_ab}. 
The average SiO abundance for the three Galactic hot cores in Table \ref{tab_X} is 1.8 $\times$ 10$^{-9}$, but their abundances range from $\sim$10$^{-11}$ to $\sim$10$^{-9}$. 

The elemental abundance of silicon in the LMC is about three times lower than the solar abundance; [Si/H]$_{\mathrm{LMC}}$ = 1.3 $\times$ 10$^{-5}$ \citep{Kor02}, while [Si/H]$_{\sun}$ = 3.6 $\times$ 10$^{-5}$ \citep{Gre98}. 
Thus the elemental abundance in the LMC partly contributes to the observed low SiO abundance, but additional processes may be necessary to account for the 1--2 orders of magnitude lower SiO abundance. 

SiO is often linked with the presence of energetic protostellar outflows since destruction of silicon-bearing dust (e.g., MgSiO$_3$) by shock is believed to be the origin of SiO \citep[e.g.,][and references therein]{Bla96,Wri96,Ter11}. 
Destruction of dust grains and release of silicon can occur around ST11 since it shows a sign of outflows as described in $\S$ \ref{sec_outflow}. 
On the other hand, the low abundance of SiO would imply that shock chemistry is not very dominant in the present chemical compositions around ST11. 
Further quantitative discussions require detailed modeling of SiO production in low metallicity environments. 
Various factors including the degree of shock, elemental abundances, and compositions of dust grains may affect the SiO abundance.

\subsubsection{CS} \label{sec_CS} 
The C$^{34}$S(7--6) transition at 337.3965 GHz ($E_u$ = 65 K) is not detected toward ST11. 
We estimate an upper limit on the abundance of carbon monosulfide (CS) to be $<$3 $\times$ 10$^{-10}$, while Galactic hot cores and high-mass protostellar objects typically show the CS abundances between 10$^{-9}$ and 10$^{-8}$ \citep[Tab. \ref{tab_X}, see also][]{vdT03}. 
CS is significantly less abundant by at least 1--2 orders of magnitude in ST11 compared to Galactic sources, and the low elemental abundances of carbon and sulfur in the LMC can not by itself explain the depletion of CS. 
Given the relatively high abundance of SO$_2$ in ST11 as discussed in $\S$ \ref{Sec_SO2}, a significant amount of gas-phase sulfur is possibly incorporated in SO$_2$ due to different circumstellar chemistry around ST11.

\subsubsection{H$_2$CS} \label{sec_H2CS} 
We detect the emission line of thioformaldehyde (H$_2$CS) at 338.0832 GHz (10$_{1,10}$--9$_{1,9}$, $E_u$ = 65 K), but the S/N of the line is relatively poor. 
The H$_2$CS abundance is estimated to be 6.2 $\times$ 10$^{-11}$ in ST11, while Galactic hot cores show the abundance of $\sim$2 $\times$ 10$^{-9}$ on average with a relatively low dispersion (Tab. \ref{tab_X}). 
H$_2$CS seems to be less abundant in the LMC hot core than in Galactic hot cores as well as CS, and only a small fraction of gas-phase sulfur is incorporated into H$_2$CS in ST11. 
Since our H$_2$CS abundance is estimated using a single line with poor S/N, further multi-line observations of H$_2$CS transitions are necessary for accurate determination of the H$_2$CS abundance in the LMC.

\subsubsection{SO} \label{sec_SO} 
The abundance of sulfur monoxide (SO) is estimated to be 2.4 $\times$ 10$^{-8}$ in ST11, while Galactic hot cores and high-mass protostellar objects typically show the SO abundances between 10$^{-9}$ and 10$^{-8}$ \citep[Tab. \ref{tab_X}, see also][]{vdT03}. 
The Orion hot core shows an exceptionally high SO abundance of 2.0 $\times$ 10$^{-7}$. 
The results suggest that ST11 shows a slightly higher abundance of SO than typical Galactic counterparts despite the low metallicity of the LMC. 
If we use the solar $^{32}$S/$^{33}$S ratio of 127, then the SO abundance in ST11 is 7.6 $\times$ 10$^{-8}$, which is even higher than typical Galactic abundances. 

We emphasize, however, that our SO abundance entails considerable uncertainty. 
The abundance is estimated using a single $^{33}$SO line at 337.1986 GHz ($E_u$ = 81 K), which contains a number of unresolved hyperfine structures, while Galactic SO abundances are estimated from $^{34}$SO lines. 
Multi-line observations of SO and its isotopologues are necessary for further discussion.

\subsubsection{SO$_2$} \label{Sec_SO2} 
Sulfur dioxide (SO$_2$) is in this study a key molecule for which we detect the largest number of transitions; we detect nine SO$_2$ lines, one SO$_2$ ($\nu$$_2$=1) line, ten $^{34}$SO$_2$ lines, and five $^{33}$SO$_2$ lines. 
The fractional abundance of SO$_2$ is estimated to be 2.1 $\times$ 10$^{-8}$ in ST11 based on rotation diagram analysis of $^{34}$SO$_2$ lines. 
The column density ratio of SO$_2$ and $^{34}$SO$_2$ is about 14, suggesting that the optically thin assumption is mostly valid for the observed SO$_2$ lines because the $^{32}$S/$^{34}$S ratio is reported to be 15 in the LMC. 
If we assume the solar isotope ratio of $^{32}$S/$^{34}$S = 22, the SO$_2$ lines could be moderately optically thick. 
In either case, we can reasonably assume that $^{34}$SO$_2$ and $^{33}$SO$_2$ lines are optically thin because these isotopologues are much less abundant than $^{32}$SO$_2$. 
The isotope abundance of $^{33}$S in ST11 based on the present results is $^{32}$S/$^{33}$S = 40, while the solar value is $^{32}$S/$^{33}$S = 127. 

Galactic hot cores typically show the SO$_2$ abundance from $\sim$10$^{-8}$ to $\sim$10$^{-7}$, while even younger embedded high-mass protostellar objects show the abundance of $\sim$10$^{-9}$ \citep{vdT03}. 
ST11 shows a factor of $\sim$3 lower SO$_2$ abundance as compared with the average abundance of 7.0 $\times$ 10$^{-8}$ for the three Galactic hot cores in Table \ref{tab_X}. 
The elemental abundance of sulfur in the LMC is reported to be [S/H]$_{\mathrm{LMC}}$ = 5.0 $\times$ 10$^{-6}$, while the solar abundance is [S/H]$_{\sun}$ = 1.9 $\times$ 10$^{-5}$ \citep{Rus92}. 
Sulfur is less abundant by a factor of $\sim$4 in the LMC compared with the solar abundance. 
Thus, the low SO$_2$ abundance in ST11 is well explained by the elemental abundance of sulfur in the LMC. 
This would suggest that hot core chemistry of SO$_2$ is dependent on elemental abundances of the host galaxy. 
A multitude of SO$_2$ and its isotopologue line detections in ST11 imply that SO$_2$ can be a key molecular tracer to test hot core chemistry in metal-poor environments. 

It should be noted that sulfur chemistry in hot core regions is highly time-dependent \citep[e.g.,][]{Cha97}. 
Hence, both the age and the interstellar environment should be taken into account to interpret the chemical compositions of sulfur-bearing species around ST11. 
Numerical simulations of hot core chemistry dedicated to low metallicity environments are thus highly required. 

\begin{figure*}[!tb]
\begin{center}
\includegraphics[width=16.2cm]{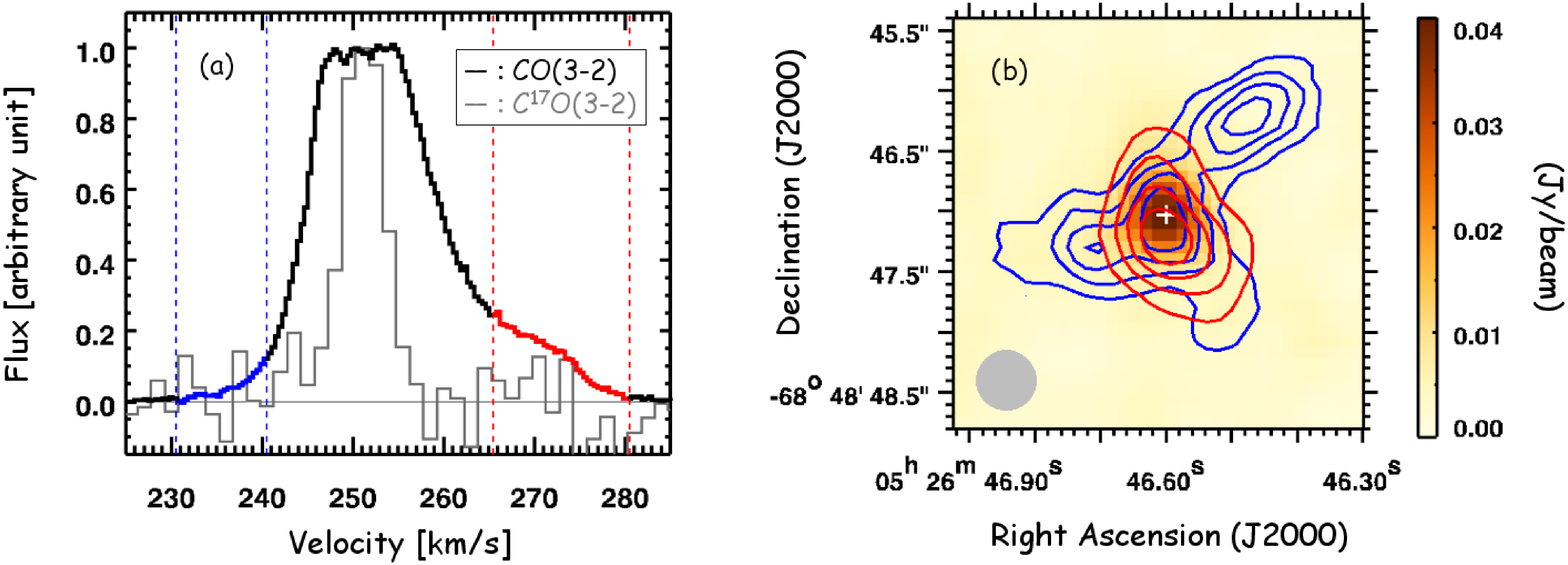}
\caption{
(a) Comparison of spectral line profiles of CO(3--2) (thick solid line, black) and C$^{17}$O(3--2) (thin solid line, gray) observed toward ST11. 
The spectra are arbitrarily scaled and the horizontal axis is the LSR velocity. 
The CO line shows a broad line width and a prominent red-shifted component, which indicate the presence of outflows in the line of sight. 
The blue and red vertical dashed lines represent the velocity range of the blue-shifted and red-shifted high-velocity components, which are visualized in the right panel. 
(b) Spatial distributions of integrated intensities of high-velocity wings. 
The blue-shifted and red-shifted wing components are shown by the blue and red contours. 
The contour levels are 20 $\%$, 40 $\%$, 60 $\%$, and 80 $\%$ of the peak intensity. 
The background is the 840 $\mu$m continuum. 
The synthesized beam size is shown by the gray filled circle at the lower left. 
}
\label{profile}
\end{center}
\end{figure*}

\subsubsection{CH$_3$OCH$_3$, HCOOCH$_3$, C$_2$H$_5$OH} \label{sec_COMs} 
Spectral lines from complex organic molecules are not detected in this work. 
We estimate upper limits on fractional abundances of three molecules whose relatively strong transitions are covered in the present data; [CH$_3$OCH$_3$/H$_2$] $<$3 $\times$ 10$^{-9}$, [HCOOCH$_3$/H$_2$] $<$2 $\times$ 10$^{-8}$, and [C$_2$H$_5$OH/H$_2$] $<$5 $\times$ 10$^{-9}$. 
Average abundances of these molecules for Galactic hot cores in Table \ref{tab_X} are [CH$_3$OCH$_3$/H$_2$] = 1.5 $\times$ 10$^{-8}$, [HCOOCH$_3$/H$_2$] = 1.8 $\times$ 10$^{-8}$, and [C$_2$H$_5$OH/H$_2$] $=$ 3.7 $\times$ 10$^{-9}$. 
These estimates suggest that CH$_3$OCH$_3$ is less abundant by at least an order of magnitude in ST11 compared with Galactic hot cores, while upper limit abundances of HCOOCH$_3$ and C$_2$H$_5$OH are comparable to the average abundances of Galactic sources. 
Although the present data do not provide conclusive upper limits on abundances of a majority of complex organic molecules in the LMC, the possibly lower abundance of CH$_3$OCH$_3$ in ST11 suggests the formation of large molecules may be less efficient in the LMC. 

It is suggested by theoretical studies that CH$_3$OH in ice mantles plays an important role in the formation of complex organic molecules \citep[e.g.,][]{NM04,Gar08a,Her09}. 
\citet{ST16} argue that formation of complex organic molecules from methanol-derived species could be less efficient in the LMC due to the low abundance CH$_3$OH ice around high-mass YSOs in the LMC. 
We confirm the significant deficiency of CH$_3$OH gas around ST11 in this work. 
We thus speculate that the low abundance of CH$_3$OH contributes to the low production efficiency of CH$_3$OCH$_3$ and possibly other complex organic molecules in the LMC. 

Note that unknown rotational temperatures and limited frequency coverages produce considerable uncertainties on the abundance estimate. 
Furthermore, the formation of large molecules from other parent species such as H$_2$CO, C$_2$H, c-C$_3$H$_2$, and NO, which are detected in the LMC, should be taken into account for comprehensive understanding of complex chemistry around protostars. 
Hence, further observations with broader frequency coverage and higher sensitivity are critically needed in conjunction with theoretical and experimental effort to understand complex organic chemistry in metal-poor environments.

\subsection{Molecular outflows} \label{sec_outflow} 
In this section, we discuss the second detection of extragalactic protostellar outflows after \citet{Fuk15}, which reported the detection of protostellar outflows in the star-forming region N159 in the LMC with ALMA. 
Evidence of protostellar outflows is seen in the CO emission line of ST11. 
Figure \ref{profile}(a) shows the spectral profile of the CO(3--2) line extracted from the 0.5$\arcsec$ diameter region centered at ST11. 
The profile of the optically thin C$^{17}$O(3--2) line is also shown for a comparison purpose. 
The CO(3--2) profile shows an apparently broader velocity width compared to other lines detected in ST11. 
The central velocity of the CO line is nearly consistent with those of other lines ($V_{LSR}$ $\sim$250), but the emission distributes in a wide velocity range from 230 km/s to 280 km/s. 
In addition, a prominent red-shifted component is seen around $V_{LSR}$ $\sim$270 km/s. 
We suggest that these high-velocity wing components are due to protostellar outflows from ST11 because the high-velocity CO gas is well spatially associated with a high-mass YSO as described in the next paragraph. 
The observed outflow velocity of $\sim$10--30 km/s is consistent with molecular outflow velocities observed in Galactic high-mass star-forming regions \citep[e.g.,][]{Lad85}. 
A high-velocity component is not obviously seen in other lines besides CO(3--2). 

Spatial distributions of high-velocity wings are shown in Figure \ref{profile}(b). 
We here define the velocity range of the blue-shifted wing to be 230.5 km/s to 240.5 km/s and the red-shifted wing to be 265.5 km/s to 285.5 km/s. 
Both high-velocity wings are spatially associated with a central protostar which is traced the the continuum emission. 
The complex structures seen in the distribution of high-velocity gas imply that the circumstellar environment of ST11 is dynamically active. 
If we assume the spatial extent of high-velocity gas to be 0.24 pc ($\sim$1$\arcsec$) and the outflow velocity to be 20 km/s according to the distribution and spectrum of the red wing component, we can roughly estimate an upper limit on the dynamical timescale of the outflow, which is about 10$^4$ years. 
This timescale is lower than a typical formation time of high-mass stars \citep[$\sim$10$^5$ years,][]{ZY07} and thus consistent with high-mass star-formation scenarios. 

We also make a rough estimate of several outflow parameters, namely, the outflow mass, the mass entrainment rate, the mechanical force, and the energy. 
The outflow mass is estimated by adding blue and red wing components within the region where outflow gas emission is detected with the S/N ratio higher than six. 
To convert integrated intensities of CO(3--2) to the total gas mass, we use a conversion factor of 8.8 M$_{\sun}$ (K km/s)$^{-1}$ pc$^{-2}$, which is derived from the present observations toward the ST11 center. 
The derived outflow mass is $M_{\mathrm{out}}$ = 74 M$_{\sun}$, where the blue-shifted component contains 13 M$_{\sun}$ ($M_{\mathrm{blue}}$) and the red-shifted component contains 61 M$_{\sun}$ ($M_{\mathrm{red}}$). 
This outflow mass corresponds to the mass entrainment rate of $\dot M = M_{\mathrm{out}} / t = $ 7 $\times$ 10$^{-3}$ M$_{\sun}$ yr$^{-1}$, where $t$ is the dynamical time scale of outflows discussed above (10$^4$ yr). 
The mechanical force ($F$) and the energy ($E$) of outflows are derived by the following equations: $F = (M_{\mathrm{blue}} V_{\mathrm{blue}} + M_{\mathrm{red}} V_{\mathrm{red}} ) / t$ and $E = (M_{\mathrm{blue}} V_{\mathrm{blue}}^2 + M_{\mathrm{red}} V_{\mathrm{red}}^2 ) / 2$. 
$V_{\mathrm{blue}}$ and $V_{\mathrm{red}}$ are mean velocities of blue and red wing components, and we use $V_{\mathrm{blue}}$ = 13 km/s and $V_{\mathrm{red}}$ = 20 km/s, respectively. 
Consequently, the derived outflow force is $F$ = 0.14 M$_{\sun}$ km/s yr$^{-1}$ and the outflow energy is $E$ = 3 $\times$ 10$^{47}$ erg. 
The above outflow parameters for ST11 are roughly consistent with those observed in Galactic high-mass YSOs that have similar luminosities with ST11 \citep[e.g.,][]{Beu02}. 

The present detection increases the number of extragalactic protostellar outflow samples, which should help understand the dynamical processes of high-mass star-formation in different metallicity environments. 
Systematic observations of extragalactic outflow sources are necessary for statistical comparison of Galactic and extragalactic protostellar outflows. 
Further detailed analysis of outflows around ST11 is beyond the scope of this paper and will be presented in a future work.

\section{Summary} \label{sec_summary} 
We report the first detection of an extragalactic hot molecular core based on radio interferometric observations toward ST11, a high-mass YSO in the LMC, with ALMA. 
The high spatial resolution (0.12 pc) ALMA Band 7 (345 GHz) spectral and continuum band observation data are presented. 
We discuss the physical and chemical properties of the source and obtained the following conclusions. 

\begin{enumerate}
\item 
Molecular emission lines of CO, C$^{17}$O, HCO$^{+}$, H$^{13}$CO$^{+}$, H$_2$CO, NO, SiO, H$_2$CS, $^{33}$SO, $^{32}$SO$_2$, $^{34}$SO$_2$, and $^{33}$SO$_2$ are detected from a compact region ($\sim$0.1 pc) associated with a high-mass YSO. 
Furthermore, a number of high excitation lines ($E_u$ $>$100 K) of SO$_2$ and its isotopologues are detected. 
On the other hand, CH$_3$OH, HNCO, CS, HC$_3$N, and complex organic molecules are not detected. 

\item 
Physical properties of ST11 are derived using the obtained data. 
The H$_2$ gas density around the source is estimated to be at least 2 $\times$ 10$^6$ cm$^{-3}$ based on the dust continuum data. 
The temperature of molecular gas is estimated to be higher than 100 K based on rotation diagram analysis of SO$_2$ and $^{34}$SO$_2$ lines. 
The SED analysis in the 1--1000 $\mu$m range suggests that ST11 is a high-mass YSO with the luminosity of 5 $\times$ 10$^5$ L$_{\sun}$ and the stellar mass of 50 M$_{\sun}$. 

\item
The compact size of the emitting source, warm gas temperature, high density, and rich molecular lines around a high-mass protostar suggest that ST11 is associated with a hot molecular core. 

\item 
We find that the molecular abundances of the hot core in the LMC are significantly different from those of Galactic hot cores. 
The abundances of CH$_3$OH, H$_2$CO, and HNCO are remarkably lower compared with Galactic sources by at least 1--3 orders of magnitude, although the gas temperature is warm enough for the sublimation of ice mantles. 
The deficiency of CH$_3$OH gas in a warm and dense region is consistent with the previously reported low abundance of the CH$_3$OH ice in the LMC.  
We suggest that the chemical compositions of ST11 are characterized by the deficiency of molecules whose formation requires the hydrogenation of CO on grain surfaces. 

\item 
It is interesting that NO shows a higher abundance in ST11 than in Galactic sources despite the notably low abundance of nitrogen in the LMC. 
This is in contrast to low abundances of nitrogen-bearing molecules such as NH$_3$, HCN, HNC in the LMC reported by previous studies. 
The reason of the enhanced abundance of NO remains to be investigated. 

\item 
Slightly lower abundance of SO$_2$ in ST11 than in Galactic hot cores is well explained by the low abundance of elemental sulfur in the LMC. 
CS and H$_2$CS are less abundant than Galactic hot cores by at least 1--2 orders of magnitude. 
The abundance of SO is possibly high in the LMC, but the estimate based on a single $^{33}$SO line should be taken with caution. 
The large number of SO$_2$ and its isotopologue line detections in the LMC hot core imply that SO$_2$ can be a key molecular species to test hot core chemistry in metal-poor environments. 

\item
We find molecular outflows around ST11, which is the second detection of an extragalactic protostellar outflow. 
An apparently broad velocity width is seen in the spectral profile of the CO(3--2) line, which ranges in the LSR velocity from 230 km/s to 280 km/s. 
A prominent high-velocity component is also seen in the red-shifted wing. 
We estimate an upper limit on the dynamical timescale of outflows to be 10$^4$ years, which is consistent with the timescale of high-mass star formation. 
Several outflow parameters are also estimated based on the present results. 

\end{enumerate}

{\footnotesize 
\acknowledgments
This paper makes use of the following ALMA data: ADS/JAO. ALMA$\#$2012.1.01108.S. ALMA is a partnership of ESO (representing its member states), NSF (USA) and NINS (Japan), together with NRC (Canada), NSC and ASIAA (Taiwan), and KASI (Republic of Korea), in cooperation with the Republic of Chile. The Joint ALMA Observatory is operated by ESO, AUI/NRAO and NAOJ. 
This work has made extensive use of the Cologne Database for Molecular Spectroscopy and the molecular database of the Jet Propulsion Laboratory. 
Partly based on data obtained at the Gemini Observatory via the time exchange program between Gemini and the Subaru Telescope (Program ID: S10B-120). 
The Gemini Observatory is operated by the Association of Universities for Research in Astronomy, Inc., under a cooperative agreement with the NSF on behalf of the Gemini partnership: the National Science Foundation (United States), the National Research Council (Canada), CONICYT (Chile), Ministerio de Ciencia, Tecnolog\'{i}a e Innovaci\'{o}n Productiva (Argentina), and Minist\'{e}rio da Ci\^{e}ncia, Tecnologia e Inova\c{c}\~{a}o (Brazil). 
The authors are grateful to Satoshi Yamamoto for his useful comment on spectral data. 
Takashi Shimonishi was supported by the ALMA Japan Research Grant of NAOJ Chile Observatory, NAOJ-ALMA-0061. 
This work is supported by a Grant-in-Aid from the Japan Society for the Promotion of Science (15K17612). 
Finally, we would like to thank an anonymous referee, whose suggestions greatly improved this paper. 
}


\end{document}